\documentclass[runningheads]{llncs}

\AtBeginDocument{%
  \providecommand\BibTeX{{%
    \normalfont B\kern-0.5em{\scshape i\kern-0.25em b}\kern-0.8em\TeX}}}

\usepackage{amsmath,amsopn,amssymb}
\usepackage{wasysym,mathtools}
\usepackage{float}
\usepackage{subcaption}
\usepackage{endnotes,microtype,xspace,graphicx,fancyvrb,multirow}
\usepackage{booktabs}
\usepackage{array,underscore,relsize}
\usepackage{enumitem}
\usepackage[labelfont=bf,font=small,skip=5pt]{caption}
\usepackage[ruled,linesnumbered]{algorithm2e}
\SetKwInput{kwInit}{Init}
\SetKwComment{Comment}{\textnormal{/*} }{ \textnormal{*/}}

\usepackage{fp}
\usepackage{siunitx}
\usepackage{hyperref}

\usepackage[acronym,nohypertypes={acronym,notation}]{glossaries}

\usepackage{tikz}
\usetikzlibrary{calc}
\usetikzlibrary{positioning}
\usetikzlibrary{fit}
\usetikzlibrary{shapes}
\usetikzlibrary{trees}
\usetikzlibrary{arrows}
\usetikzlibrary{overlay-beamer-styles}

\newcommand{\sys}{\mbox{\textsc{S2malloc}}\xspace}
\newcommand{\centered}[1]{\begin{tabular}{l} #1 \end{tabular}}

\newcommand*{\circledbullet}{%
    \ooalign{$\circledcirc$\cr\hidewidth$\bullet$\hidewidth}%
}

\newsavebox{\fmbox}

\usepackage{verbatimbox}

\newcommand{\cc}[1]{\mbox{\smaller[0.5]\texttt{#1}}}

\input{code/fmt}

\def\Snospace~{\S{}}

\usepackage{pifont}% http://ctan.org/pkg/pifont

\if 0

\fi

\newif\ifdraft\drafttrue
\newif\ifnotes\notestrue
\ifdraft\else\notesfalse\fi

\input{glyphtounicode}
\newcommand{\squishlist}{
\begin{itemize}[noitemsep,nolistsep]
  \setlength{\itemsep}{-0pt}
}
\newcommand{\squishend}{
  \end{itemize}
}

\usepackage{tikz}
\newcommand*\WC[1]{%
\begin{tikzpicture}[baseline=(C.base)]
\node[draw,circle,inner sep=0.2pt](C) {#1};
\end{tikzpicture}}

\newcommand*\BC[1]{%
\begin{tikzpicture}[baseline=(C.base)]
\node[draw,circle,fill=black,inner sep=0.2pt](C) {\textcolor{white}{#1}};
\end{tikzpicture}}

\usepackage{xstring}
\newcommand{\PP}[1]{
\vspace{2px}
\noindent{\bf \IfEndWith{#1}{.}{#1}{#1.}}
}

\newcommand{\leftFigure}[1]{
\centered{\vspace{-0.3cm}\includegraphics[width=.35\textwidth]{#1}}
}

\newcommand{\rightFigure}[1]{
\centered{\vspace{-0.3cm}\includegraphics[width=.35\textwidth]{#1}} 
}

\newcommand{\boxbeg}{
\vspace{2px}
\noindent\begin{tabular}{|l|}\hline
\begin{minipage}{3.2in}
\vspace{2px}
\noindent
}

\newcommand{\boxend}{
\vspace{2px}
\end{minipage}\\ \hline
\end{tabular}
\vspace{-10pt}
}

\usepackage{url}

\usepackage{ifthen}
\newboolean{showtext}
\setboolean{showtext}{false} % Set to false to hide the text

\newcommand{\rt}[1]{%
  \ifthenelse{\boolean{showtext}}{\textcolor{orange}{#1}}{%
  }%
}

\begin{document}

\title{\sys: Statistically Secure Allocator for Use-After-Free Protection And More}

\author{Ruizhe Wang\orcidID{0009-0001-5607-3917}\and
Meng Xu\orcidID{0009-0001-6364-4837} \and \\
N. Asokan\orcidID{0000-0002-5093-9871}}
\authorrunning{R. Wang et al.}
\institute{University of Waterloo \\
\email{\{ruizhe.wang, meng.xu.cs\}@uwaterloo.ca asokan@acm.org}}

\date{}

\maketitle

\begin{abstract}
    Attacks on heap memory, encompassing
    memory overflow,
    double and invalid free,
    use-after-free (UAF),
    and
    various heap spraying techniques
    are ever-increasing.
    Existing entropy-based secure
    memory allocators provide statistical defenses
    against virtually all of these attack vectors.
    Although they claim protections against UAF attacks,
    their designs are not tailored to \emph{detect}
    (failed) attempts.
    Consequently,
    to beat this entropy-based protection,
    an attacker can simply launch the same attack repeatedly
    with the potential use of heap spraying
    to further improve their chance of success.

    %In response to this shortfall, 
    We introduce \sys,
    aiming to enhance UAF-attempt detection
    without compromising other security guarantees
    or introducing significant performance overhead.
    To achieve this,
    we use three innovative constructs in secure allocator design:
    \textbf{free block canaries (FBC)} to detect UAF attempts,
    \textbf{random in-block offset (RIO)}
    to stop the attacker from
    accurately overwriting the victim object, and
    \textbf{random bag layout (RBL)} to impede
    attackers from
    estimating the block size based on its address.

    We show that
    (a) by reserving 25\% of the object size for the RIO offset,
    an 8-byte canary offers a 69\% protection rate if the attacker
    reuses the same pointer and 96\% protection rate if the attacker does not, 
    against UAF exploitation attempts targeting a 64 bytes object,
    with equal or higher security guarantees against all other attacks;
    and
    (b) \sys is practical, with only a 
    2.8\% run-time overhead on PARSEC and an 11.5\% overhead on SPEC.
    Compared to state-of-the-art entropy-based allocators,
    \sys improves UAF-protection
    without incurring additional performance overhead.
    Compared to UAF-mitigating allocators,
    \sys trades off a minuscule probability of failed protection
    for significantly lower overhead.

    \vspace{-5pt}
\end{abstract}

\section{Introduction}
\label{s:intro}

Heap-related vulnerabilities are serious and common threats
that can be leveraged to launch attacks resulting in 
arbitrary code execution or information leakage.
Heap overflow, double and invalid free,
and use-after-free (UAF) are among the most
common types of these vulnerabilities.
According to the Common Vulnerabilities and Exposures (CVEs)
report of 2022, they were ranked 19th, 11th, and 7th respectively,
in terms of bug prevalence~\cite{cwetop25}.

Secure memory allocators are
an important defense
against heap exploitations.
State-of-the-art secure allocators
can be generally classified into two categories:
\emph{entropy-based} generic memory allocators and
 \emph{UAF-mitigating} allocators.

UAF occurs when a previously freed memory block
is used to store data.
It receives special attention in secure allocator design
due to
its prevalence and
the powerful exploitation primitives
(e.g. arbitrary read/write) it enables.
Chromium has reported that 
more than one-third of their security issues are related to UAF,
more prevalent than other types of memory errors combined~\cite{chromium-report}.
While complete mitigation of UAF is theoretically possible
by virtually never re-using a previously freed memory block,
UAF-mitigating allocators incur 
substantial overheads (e.g., about 40\% for MarkUs~\cite{markus}),
and their complexity leaves them vulnerable
to new attacks~\cite{ffmallo} (albeit preventable).

On the other hand,
entropy-based heap allocators
aim to provide comprehensive protection against
most, if not all, common heap vulnerabilities with
simpler designs
but may fail with a small probability.
Specifically
on UAF mitigation,
entropy-based memory allocators typically use
delayed free-lists~\cite{guarder,slimguard} to prevent the same memory block
from being \emph{immediately} re-allocated
after being freed.
Attackers now face a moving-target even when
they manage to obtain a dangling pointer as 
they have less confidence in knowing 
when this pointer becomes valid again and/or
which object it might point to.
While achieving relatively low overhead on
time usage,
especially compared with UAF-mitigating allocators
(see~\autoref{s:bg} for details),
existing entropy-based allocators
still face the challenge of entropy-loss due
to heap spraying, information leak, and silent 
failures on (potentially repeated) trials.

To overcome these challenges,
we propose \sys,
a novel heap allocator that
combines the best of both types of 
memory allocators---high
security assurance against not only UAF but also
other types of heap vulnerabilities
and yet with low memory and computation overhead
 on par with state-of-the-art
entropy-based memory allocators.
\sys achieves its promises
through several innovative constructs:
\textbf{randomized in-slot offset (RIO)},
\textbf{free-block canary (FBC)}, and
\textbf{random bag layout (RBL)}.
RIO mitigates UAF attacks by allocating blocks with random offsets,
obstructing the attacker from locating the target field in a data structure.
FBC puts cryptographically secure canaries in free blocks to
detect illegal writes,
turning a failed UAF exploitation attempt into a clear signal.
RBL organizes blocks of the same size range using sub-bags.
Only blocks within the same memory page are guaranteed to be in the same sub-bag.

\PP{Summary}
We claim the following contributions: we
\begin{itemize} [noitemsep, topsep=0pt,leftmargin=12pt]
    \item analyze current entropy-based allocators in
          real-world UAF attack scenarios and show that 
          the level of protection they provide is not as strong as 
          originally claimed~(\autoref{s:motivation}).
    \item  present \sys, a straightforward drop-in solution 
          addressing the above weaknesses,
          while also protecting against all other commonly observed heap memory vulnerabilities.
          \sys does not require any special hardware,
          program recompilation, or
          elevated privileges,
          and works on both x86 and AARCH,
          thus increasing its deployability~(\autoref{s:design}). 
    \item through various real-world CVEs and benchmarks,
           show that \sys can successfully detect all attacks
          while incurring a 2.8\% execution delay and 27\% memory overhead 
          on the popular PARSEC benchmark and 11.5\% and 37\% respectively
          on SPEC~(\autoref{s:formal},~\autoref{s:eval}).  
\end{itemize}
    The
    \href{https://github.com/ssg-research/s2malloc}{software artifact} is open-sourced.

\section{Background and Related Work}
\label{s:bg}

In this section,
we give a succinct overview of
common heap vulnerabilities and the 
most recent advancements in two defense lines:
moving-target defense against generic heap exploitations
and complete mitigation against use-after-free (UAF)---the current
most frequently encountered type of memory error~\cite{chromium-report}.

\subsection{Heap vulnerabilities}
\label{ss:bg-vulns}

Common heap vulnerabilities include
\BC{1}
buffer overflow
(reads/writes to an out-of-bound memory location),
\BC{2}
double free
(frees an already-freed object),
\BC{3}
invalid free
(frees an invalid pointer),
and,
as explained later in details,
\BC{4}
use-after-free (UAF).
Successful exploitation of these vulnerabilities
can cause heap corruption, leading to
devastating results such as
denial of service (DoS),
information leak,
arbitrary code execution, and/or
privilege escalation.

\PP{Allocator-assisted UAF}
Use-after-free (UAF) is a common heap error that occurs when
a program unintentionally releases a heap object
but continues to use the original pointer,
i.e., a dangling pointer.
Abusing the dangling pointer,
an attacker can create powerful exploitation primitives
such as arbitrary-address-write and code execution.
A typical UAF exploitation involves the following steps:
\begin{itemize} [noitemsep, topsep=0pt,leftmargin=12pt]
    \item Trigger the vulnerable \cc{free()} function to release a
          heap object $A$,
          turning the pointer $P$ that points to
          the freed object into a dangling pointer.

    \item Look for a victim heap object $B$
          having ideally 
          the same size as the released object.
          Victims containing interesting data such as pointers or length
          are usually preferred.

    \item Wait for the program to allocate
          a new $B$ object
          that could refill the memory space
          initially belonging to $A$.
          The attacker can now
          corrupt the victim object $B$ with pointer $P$.
          This primitive can be used to
          either hijack the control flow or
          corrupt sensitive
          data such as \cc{uid} or \cc{gid}.
\end{itemize}

\subsection{Entropy-based allocators}
\label{ss:bg-entropy-alloc}

Entropy-based heap allocators strive to
provide protections against
almost all of the above-mentioned heap exploitations
by minimizing the attacker's success rate.
Ideally, the success rate should be low enough
to deter attackers from even trying to attack the system.
However, practical implementations face limitations 
in terms of memory and computation resources,
as demonstrated in Guarder~\cite{guarder}.

\PP{BIBOP}
State-of-the-art entropy-based allocators typically
build on the \emph{Big Bag of Pages} (BIBOP)~\cite{bibop}
memory management mechanism
with various security enhancements.
BIBOP-style allocators classify allocation sizes as classes.
For each size class,
one or several continuous pages are treated as a bag,
and all allocations of the same size class
will be assigned to a corresponding bag.
Each bag is split into several slots preemptively and
each allocated object will take one of them.
The status of each slot is monitored by a bitmap and
can be used to defend against double or invalid frees.
Bag metadata is stored separately to
avoid metadata-based attacks~\cite{how2heap},
and UAF only occurs within the same size class.

\PP{Extended secure features}
Prior works have introduced
a diverse set of security enhancements
over the basic BIBOP-style allocation,
including:

\begin{itemize} [noitemsep, topsep=0pt,leftmargin=12pt]
    \item \emph{Guard page}:
          is a single-page virtual memory block
          not mapped to the physical memory.
          Therefore,
          any attempt to dereference an address in a guard page
          triggers a segmentation fault.
          Guard pages could be strategically placed after each bag
          or randomly within bags
          to protect against overflows or random accesses.

    \item \emph{Heap canary}:
          is a small object set to a secure value and
          placed at the end of each slot.
          Heap overflow can be detected if the canary value is modified.

    \item \emph{Random allocation}:
          guarantees that slots within each bag
          are not allocated sequentially (i.e., linear allocations).
          Instead, each allocated slot is randomly chosen from
          at least $r$ free slots.
          More slots will be requested if there are not enough
          free slots to satisfy this requirement.
          Intuitively, the entropy (reflected by $r$)
          marks a trade-off between security and performance overhead.
\end{itemize}

\PP{Evolution history}
While many entropy-based allocators have been proposed,
we introduce three representative works
that have contributed to advancing the field. 

\emph{DieHarder}~\cite{novark2010dieharder}
is one of the earliest entropy-based secure allocators
that adopts the BIBOP-style memory management and
provides statistical protections against
heap exploitations.
Despite incorporating all the security features mentioned above, 
DieHarder has several issues compared to more recent works.
These include
unstable allocation entropy,
predictable guard pages,
and a significant impact on
the execution time of the hardened program.

\emph{Guarder}~\cite{guarder}
offers multiple improvements over DieHarder, such as
using a linked-list-based free block management algorithm
instead of bitmap traversal,
providing stable allocation entropy,
and offering in-bag tunable random guard pages.
Guarder significantly
reduces the run-time overhead to less than 3\%, 
making it suitable for production systems.

\emph{SlimGuard}~\cite{slimguard}
further extends Guarder by
reducing its memory overhead.
It divides size classes into 16-byte increments
instead of powers of two.
However, the current
implementation of SlimGuard
has two security compromises: it
1) allocates blocks sequentially
in the free-list, violating the claim of 
random allocation.
2) reuses freed blocks to store metadata,
possibly due to an implementation flaw, 
violating the argued metadata separation practice.

\subsection{UAF-mitigating allocators}
\label{ss:bg-uaf-alloc}

\rt{Conventional heap allocators
prioritize defragmentation and memory saving
and hence,
tend to eagerly re-allocate an object
to a recently freed memory block of a similar size.
An attacker can easily abuse this allocation strategy
to
predict future allocations.
In other words,
this defragmentation strategy favors the attackers and 
makes UAF exploitation (\autoref{ss:bg-vulns}) easier.
The eager reallocation strategy is based on implicitly 
trusting that the \cc{free()} operation issued by the developer
is an accurate indication of
the end of the lifetime of the memory object.
In reality, however,
people make mistakes and a \cc{free()} call
might be issued when there are still
dangling references.}

UAF-mitigating allocators
specialize in UAF-protection only and
can be broadly categorized into the following types:

\begin{itemize} [noitemsep, topsep=0pt,leftmargin=12pt]
    \item[A] \emph{Pointer validation on dereference}~\cite{dangsan,ptauth,FreeSentry}: 
    i.e., checks whether a pointer is valid and safe
    for a read/write operation upon dereference.
        
    \item[B] \emph{Pointer invalidation on free}~\cite{dangzero,oscar,Mpchecker,corncopia,ffmallo}: 
    i.e., whenever \cc{free} is called on a pointer,
    it revokes its capability to access the associated virtual address.

    \item[C] \emph{Memory sweeping}~\cite{lee:dangnull,markus,minesweep}: 
    i.e., checks all pointers stored in memory and
    either actively removes all dangling pointers or 
    reuses a freed memory chunk with the assurance that
    no dangling pointer to this freed chunk exists.
    
    \item[D] \emph{Context-based allocation}~\cite{cling,typeaftertype,dhurjati2003memory}:
    i.e., permits the re-allocation of freed memory chunks
    only to objects allocated in the same ``context.''
\end{itemize}

\noindent
Not all UAF-mitigating allocators are drop-in replacements
of the system memory allocator,
as some of them require
recompilation of the protected programs~\cite{cling,typeaftertype,wang2024semalloc},
special hardware~\cite{corncopia}, or
kernel modifications~\cite{dangzero}.

\PP{Security guarantee} 
UAF-mitigating allocators in categories A, B, and C
can eliminate all UAF exploits (hence complete mitigation),
although some of them
might incur large overheads or
require customized kernel or hardware.
Context-based allocators (category D),
first proposed in~\cite{dhurjati2003memory} as a heap-safety property,
typically run faster
but offer incomplete protection.
For example,
Cling~\cite{cling} defines the allocation context
based on the two innermost return addresses on the call stack
when \cc{malloc} is invoked.
Suppose two memory allocations occur in the same context;
in this case,
the object allocated the second time
may occupy the same memory chunk allocated the first time
(if the first object is freed).
TypeAfterType~\cite{typeaftertype} defines the allocation context
based on object types
(e.g., the type \cc{int} passed in \cc{malloc(sizeof(int))},
and objects may be reallocated on freed memory chunks
originating from the same allocation context.
In summary,
UAF is still possible in context-based allocators.

However, unlike entropy-based allocators (including \sys), which also mitigate UAF imperfectly, the protection from context-based allocators is deterministic.
Specifically,
if the dangling pointer and the target object
are allocated in different contexts,
there is zero chance of success in corrupting the target object,
regardless of the attackers' strategies.
On the other hand,
context-based allocators provide no protection
if the dangling pointer and the target object
fall in the same context.
Entropy-based allocators provide probabilistic protection
in both cases.
We will discuss the implications
via CVEs in~\autoref{sec:real-world-example}.

\PP{Porting for UAF-write defense only}
Allocators that instrument runtime checks for pointer validity
(i.e., category A) can opt to
trade protections of UAF-read attacks for performance
by checking the pointer validity on write accesses only.
In fact,
as reported in~\cite{singh2019},
only 5\% to 25\% of memory accesses are write accesses
in the SPEC 2017 benchmark,
indicating a potential reduction in overhead
for category-A allocators.
On the other hand,
for allocators in categories B, C, and D,
splitting UAF-read and UAF-write protection is inherently hard
as their designs do not differentiate between read and write accesses.

\section{Motivation}
\label{s:motivation}

While effective in defending against
various heap exploits,
entropy-based allocators are not
ideally suited to protect against UAF attacks.
This partially motivates the
stream of research on UAF-mitigating allocators
presented in~\autoref{ss:bg-uaf-alloc}.
Specifically,
there exist easy-to-identify
blind spots that
drastically reduce the efficacy
of defending UAF using
random allocation
and increase attackers' confidence on
launching a successful attack without trying to
predict the next allocation slot.
In this section,
we first present the adversary model
followed by a discussion on the
limitations of existing entropy-based allocators,
which motivates the development of \sys.

\subsection{Adversary model}
\label{ss:design-adv-model}

We assume that the attacker can analyze
the source code and binary executable
to determine the implementation details of the victim program,
including vulnerabilities and other relevant information
such as the size and layout of critical data structures.
We also assume that the attacker can identify
when a victim object is allocated or de-allocated.

We do, however, assume that
the underlying kernel and hardware are trusted and
an attacker cannot utilize a data leakage channel,
like \cc{/proc/\$pid/maps},
to discover the location of the heap allocator's metadata.
The attacker cannot compromise the random number generator
nor can they take control of the heap allocator.
Exploiting bugs of the allocator itself is out of scope.
These assumptions are similar to that of
other entropy-based allocators~\cite{guarder,slimguard}.

Additionally, we allow
attackers to use any existing heap feng-shui~\cite{heap-feng-shui} technique (e.g., heap spray)
to prepare or manipulate the layout of heap to facilitate UAF-exploits.
And attackers can retry an exploit
as long as previous attempts fail silently.
These assumptions make our adversary model
stronger than those assumed in
entropy-based allocators and on-par with
the adversary models in
UAF-mitigating allocators~\cite{markus,ffmallo}.

\subsection{Challenge 1: entropy loss}
\label{sec:motivation-example}

Entropy-based allocators thwart UAF
by avoiding instant memory reuse.
However, if
1) the attacker could continue to retry the attack
when the previous trial fails, or
2) the heap memory can be spoofed with
either dangling pointers or victim objects,
it is guaranteed that the attack would eventually succeed
even without the victim's notice.

\autoref{code:uaf-example} is an example,
abstracted from mRuby issue 4001~\cite{mruby-4001},
a UAF vulnerability in the Ruby compiler.
The function \cc{mrb_io_initialize_copy} is called when opening a file. 
It first frees the existing data pointer of the \cc{copy} object (\cc{DATA_PTR(copy)}) (line 7) and allocates new memory to it (line 9 and 11). 
If an invalid argument is passed, calling \cc{io_get_open_fptr} would throw an exception (line 10), making \cc{DATA_PTR(copy)} a dangling pointer to the freed object.

Using this vulnerability, the attacker can allocate a string object to take the freed memory space.
The attacker can then close the file in Ruby, which will set the first word of the pointed memory to \cc{INF}. If this memory is taken by a string object, its length will be overwritten to \cc{INF} 
that allows arbitrary memory read and write.

\begin{figure}[h]
    \centering
    \hfill\begin{minipage}[c]{.9\linewidth}\input{code/uaf-motivating-example}\end{minipage}
    \caption{Example UAF attack based on mRuby issue 4001~\cite{mruby-4001}}
    \label{code:uaf-example}
\end{figure}

In the above scenario,
random (i.e., non-sequential) allocation or
delayed free-lists
available in existing
entropy-based allocators~\cite{novark2010dieharder,guarder,slimguard}
merely increases the attack difficulty:
as long as the attacker can wait,
the memory chunk referred to by the dangling pointer will eventually be re-allocated,
allowing the UAF exploit to proceed after some delay.

Furthermore,
the entropy diminishes if an attacker
is allowed to repeat the same attack
\emph{without penalty}
(e.g., when a failed attempt does not
crash the target program or trigger attention).
Similarly,
if attackers have the ability of
spraying the heap with
either dangling pointers of victim objects,
the probability of success increases significantly.

This motivates us to design an allocator in a way that
1) actively searches for UAF attempts
and raises signals if the evidence is found; and
2) stops the attacker from
locating critical information in memory even if
the attacker manages to obtain a dangling pointer.
In this example,
we could prevent the attacker from
being able to deterministically locate the string length
even if the attacker manages to obtain a dangling pointer
to a string object originally pointed to by \cc{DATA\_PTR(copy)}.
In addition,
any attempts of writing to unallocated memory
will be detected with a high probability.
If an attacker attempts to spray the heap
with arbitrary write to increase success rates,
we can raise a signal on or even before the
UAF actually happens.

\subsection{Challenge 2: information leak}
\label{sec:info-leak}

Existing entropy-based allocators~\cite{guarder,slimguard}
create a huge memory pool for each block size range,
resulting in the leakage of block size via their address,
possibly revealing the victim software's internal state to the attacker.
For example,
each BIBOP bag is assigned an 8 Gigabytes virtual memory pool
in SlimGuard~\cite{slimguard}.
Any objects belonging to this bag will be allocated from this pool.
As a result,
obtaining a known sized block will be sufficient enough
to infer the size of any blocks sharing the upper 32-bit address.
Further,
block addresses in Guarder~\cite{guarder} are aligned by their size.
Block size would possibly be inferred just based on its address.
For example,
a block with an address value ending with
\cc{0x10} is of size 1 to 16 bytes,
and a block with an address value ending with
\cc{0x300} is highly likely of size 129 to 256 bytes.

We mitigate this threat by dividing bags into sub-bags,
limiting the size leakage only if the attacker-controlled block
resides in the same sub-bag as the victim block.
Furthermore, we assign random guard pages within sub-bags to make the sub-bag boundaries unpredictable.

\section{Design and Implementation}
\label{s:design}

Now
we explain the design and implementation of \sys and
 how it thwarts the types of heap attacks
 in~\autoref{ss:bg-vulns}.

\subsection{Architectural overview}
\begin{figure*}
    \centering
    \includegraphics[width=\textwidth]{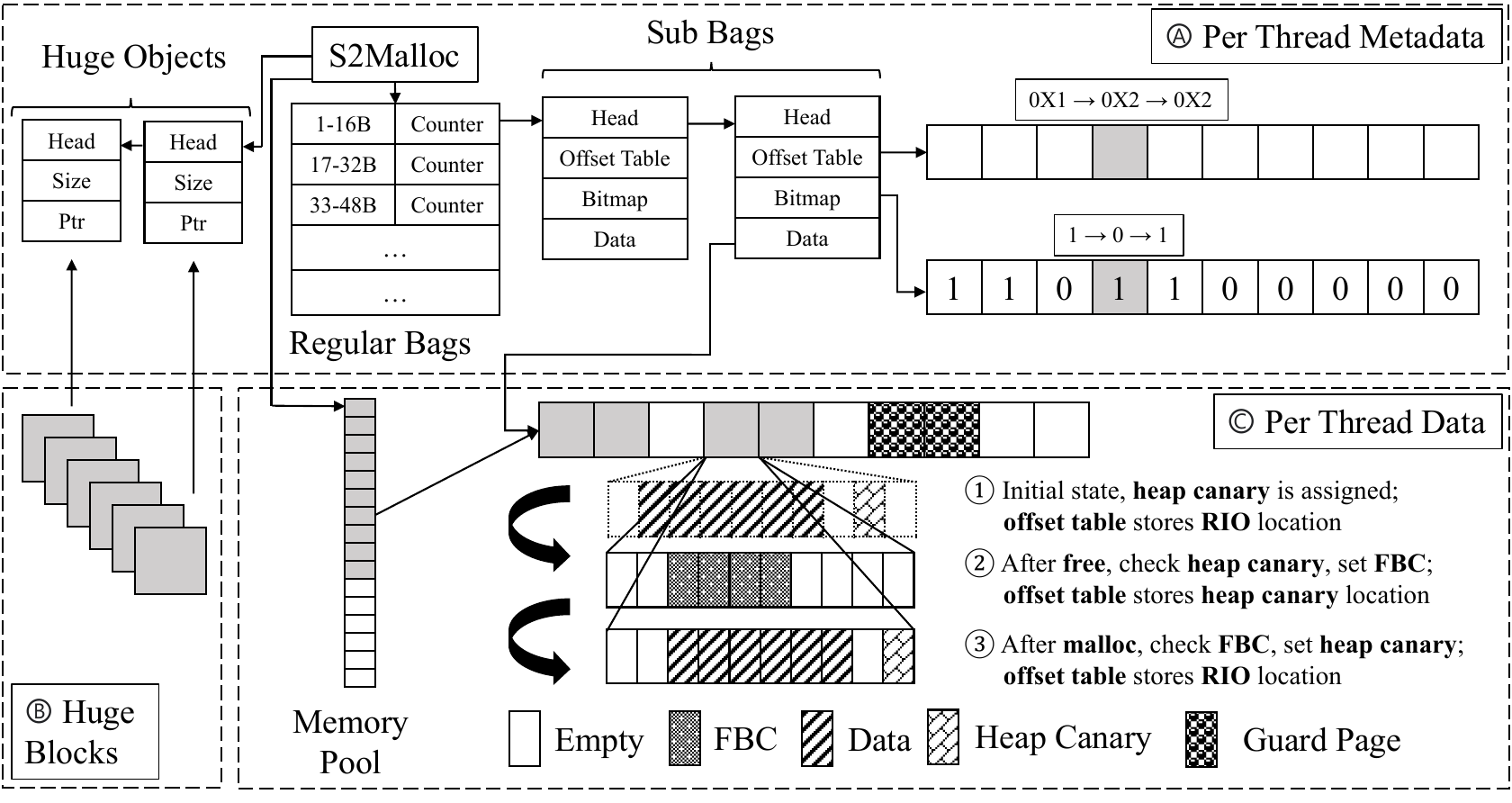}
    \caption{Overview of \sys with an example of free and malloc. \protect\WC{A}, \protect\WC{B}, and \protect\WC{C} show three \sys segments, stored in segregated memory. \protect\WC{1}, \protect\WC{2}, and \protect\WC{3} show how an allocated bag slot is freed and then allocated.}
    \label{fig:block-remalloc}
\end{figure*}

At its core,
\sys adopts BIBOP to manage memory blocks.
An overview of \sys is illustrated in~\autoref{fig:block-remalloc}.
\sys maintains a per-thread metadata (\WC{A}), 
stored in a memory chunk requested directly from the kernel.
Huge blocks are obtained or released from the OS directly (\WC{B}),
and are stored using a linked list.
Small objects are maintained using bags,
claiming memory indirectly from a segregated memory pool (\WC{C}).
Each \textbf{regular bag} maintains the metadata of blocks of a size range,
including the number of free slots and a list of \textbf{sub bags}. 
We take \textbf{sub bag} as the basic unit of a group of slots
~(\autoref{sec:random-bag-layout}).

The data field points to a memory chunk requested from the memory pool
to store the objects allocated to this sub-bag.
The bitmap indicates whether the current slot is taken or not.
If the current slot is taken,
the corresponding offset table cell stores an offset indicating
where the data stored in the bag starts (\autoref{sec:rio}).
Otherwise, the slot is free and
the offset table stores the location of the FBC (\autoref{sec:fbc}).

When a \cc{free()} call is received by \sys
(step \WC{1} $\rightarrow$ \WC{2}),
\sys checks whether:
(1) the bitmap indicates the current block is taken;
(2) the offset stored in the offset table matches with the freed pointer address;
and (3) the canary value is not modified.
If all checks pass,
the current block will be freed:
the bitmap cell will be set to free and
an FBC will be put at a random location in the current block.
The offset table will be updated to store the location of the FBC.

When a \cc{malloc()} call comes,
(step \WC{2} $\rightarrow$ \WC{3}),
\sys randomly selects one free block of the corresponding size and
checks the FBC of current and nearby free blocks.
A random offset will be generated indicating
where the data starts within the current block,
and the offset table will be updated accordingly.
The heap canary will be set after the last data byte of the current block,
and the bitmap will be updated.
In this example,
we assume that each block stores at most 7 bytes of data.
The heap canary is set to the 9th byte initially as the offset is one
and is then set to the 10th after reallocation as the offset is changed to two.

\subsection{Randomized in-slot offset (RIO)}
\label{sec:rio}

In all existing secure memory allocators,
allocated objects store their data from the first byte of the allocated slot.
Alternatively,
we propose that the object will be stored with a random offset $p$,
and the first $p$ bytes of the slot will be left empty.
After the slot is freed and allocated to another object,
the offset $p$ will be re-computed.
Thus, the relative offset between these two objects
cannot be accurately predicted and
the attacker cannot accurately re-use a freed pointer
and arbitrarily read or write the target memory.

We define $k$ to be the RIO entropy.
For each bag with blocks of $b$ bytes,
$e = b / k$ bytes are not used to take data,
and each block can take at most $b - e$ bytes of data,
guarantee minimal in-slot offset entropy.
We refer to these extra bytes as entropy bytes.
Suppose this block is \cc{malloc}-ed with an object of $s$ bytes ($s < b - e$).
Before this block is allocated,
$p \in (0, b - s)$ is computed
to decide the starting byte of the data object.
$p$ is 16-byte aligned
following the minimum default alignment of
GNU C implementation~\cite{glibcalignment},
and to be compatible with special data structures,
such as atomic objects that need to be stored to align with the 
registers and cache lines.
The offset of each block will be stored separately in an offset table.
The minimal entropy $e$ increases as $b$ becomes larger
to avoid introducing high memory overhead for small objects and
to provide stronger protections for large objects,
observing the fact that
larger objects have more complicated structures and
are more likely to be targeted.

We adopt the Permuted Congruential Generator (PCG)~\cite{pcg2014} algorithm following the design of ~\cite{slimguard}
to generate all random numbers.
With negligible execution time -- much faster than existing random number generators,
such as LCG and Unix XorShift, 
PCG generates hard-to-predict numbers.
While it does not provide cryptographic security, 
the only known attack towards PCG requires three consecutive PCG
outputs to recover the seed~\cite{pcg-attack}.
However, in \sys, none of the generated numbers is accessible to the user and can only be
obtained by sweeping the memory.
As a result, PCG is sufficient enough to provide the required level of entropy.

\subsection{Random bag layout (RBL)}
\label{sec:random-bag-layout}

As with existing secure allocators,
\sys employs BIBOP-style management for small-size blocks.
Blocks larger than 64 kilobytes
are mapped and unmapped from the OS directly,
and are managed using a linked list.
Blocks smaller than 64 kilobytes are further classified as
small, medium, and large blocks
to decrease the number of size classes.
Small bags contain blocks smaller or equal to 1 kilobyte,
and a bag is created every 16 bytes (16 bytes granularity).
For example,
the first small bag takes blocks smaller than 16 bytes,
and the second small bag takes blocks of (16, 32] bytes
(without taking RIO into consideration).
Medium bags contain blocks within the range of 1 kilobyte and 8 kilobytes
with the granularity of 512 bytes;
large bags contain blocks within the range of 8 kilobytes and 64 kilobytes
with the granularity of 4 kilobytes.
In total, \sys has
64 small bags,
14 medium bags, and
14 large bags.

\sys obfuscates the virtual memory allocation and
stops linking block sizes to their addresses.
Instead of allocating a dedicated virtual memory pool for each bag,
(as shown in prior works~\cite{guarder,slimguard}),
we create a single virtual memory address pool for all bags.
We further divide each bag into sub-bags each containing 256 slots.
Each bag creates new sub-bags upon need,
and a newly created sub-bag would request corresponding memory from the pool.
We use a bump pointer to track the available memory in the pool and
linearly allocate pool memory to sub-bags.

\sys randomly places guard pages within sub-bags
to thwart overflow, spraying, and random pointer access.
If a sub-bag is randomly allocated with a guard page,
one of its pages will be unmapped (protected) randomly
using the \cc{mprotect} system call.
Any slots within this protected page
will be marked as allocated in the bitmap
to avoid allocating them to the program.
Any accesses to these slots are thus invalid and result in a segmentation fault.
The tunable guard page rate can be configured in an environment variable.
We note that the memory pool allocation is not deterministic and
cannot be predicted due to the random guard pages.

\sys guarantees that the block size leakage occurs
only if the known block and the victim block reside on the same memory page:
adjacent blocks may not be within the same sub-bag,
and RIO guarantees that
blocks start at addresses that cannot be deterministically predicted.
On the contrary, two blocks are highly likely of the same size range in SlimGuard
if their address difference is smaller than 8GB.

\begin{table*}[t]
    \centering
    \small
    \begin{tabular}{c|cccc|c|cc|cc}
        \toprule
        Secure & Guard & Rand. & Segre. & Heap & Ptr & UAF & UAF & \multicolumn{2}{c}{Overheads} \\
        Allocators & Pages & Alloc. & MD & Canary & Inval. & Mitigate & Detect & Memory & Runtime                                         \\
        \midrule
        DieHarder & Y                                                               & Y          & Y          & N          & N          & \LEFTcircle              & N         & 21.3\%          & 2.1\%          \\
        Guarde               & Y                                                               & Y          & Y          & Y          & N          & \LEFTcircle              & N         & 58.1\%          & 2.4\%          \\
        SlimGuard         & Y                                                               & N          & N          & Y          & N          & \LEFTcircle             & N         & 22.5\%          & 4.4\%          \\

        \midrule
        \sys                        & Y                                                     & Y & Y & Y & N & \CIRCLE & High prob. & 26.8\%& 2.8\% \\        
        
        \midrule
        MarkUs             & N                                                               & N          & N          & N          & Y          & \CIRCLE & N    & 13.0\%$^\ast$            & 42.9\%$^\ast$         \\
        FFmalloc              & N                                                               & N          & N          & N          & Y          & \CIRCLE & Y        & 50.5\%$^\ast$          & 33.1\%$^\ast$         \\
        \bottomrule
    \end{tabular}
    \caption{
        Overview of existing secure memory allocators and \sys to illustrate how \sys fills the gap (MD: metadata, \LEFTcircle: One-shot attack only, \CIRCLE: Repeated/spray attacks).
        Overheads are measured on PARSEC~\cite{parsec}, detailed in~\autoref{s:eval}.
        Overheads of MarkUs and FFmalloc
        (numbers marked with $\ast$)
        are reported in~\cite{ffmallo} instead of measured by us.
        \protect\rt{With that said,
        the performance numbers
        shown here are for a qualitative illustration
        on the scale of overhead only.
        For quantitative comparisons,
        please refer to details in~\autoref{s:eval}.}
    }
    \label{tab:allocators}
\end{table*}

\subsection{Hardening heap canaries}
\label{sec:regular-canary}

Canary is a small data block
put after the allocated memory to detect overflow,
initially introduced in StackGuard~\cite{stackguard} to protect the stack.
Canary has now been adopted to protect the heap~\cite{heapsentry}.
At the time a memory slot is being allocated,
the canary will be set to a specific value.
This value will be checked at the time this slot is freed,
and memory overflow will be detected if the canary value changes.

However, in previous designs,
this value is set to be either
globally identical~\cite{guarder} or
is binded with slots~\cite{slimguard},
and could be trivially broken by a knowledgeable attacker.
We follow the design of previous works to use 
the secure MAC of the memory address as the canary~\cite{pecan,ccfi}.
Specifically,
we take the CMAC-AES-128 encrypted block address
as the canary implemented using AES-NI~\cite{aesni} (on x86 CPUs) or
Neon~\cite{neon} (on AARCH CPUs) to
keep the canary confidential and compact.
Even if the attacker learns a canary value, they can only use it to
break the current object
or any further objects allocated to this slot with the same RIO.

Specifically in \sys,
we put a $\iota$-byte canary immediately after the last 
data-storage byte in the allocated slot,
(i.e., the $(p + b - e)$th byte),
and the canary will be checked upon free.

\subsection{Free block canaries (FBC)}
\label{sec:fbc}

Existing entropy-based allocators defend against UAF-write attacks
by statistically avoiding allocating 
a victim object in a block pointed to by a dangling pointer.
Although a failed attack attempt only modifies a free block
without causing any harm,
the attack attempt is not detectable either and
given the fact that the same attack can be retried,
the attacker will succeed eventually.

To detect such attempts,
we put a canary of length $c$ in each free block.
The canary value is also computed using CMAC-AES-128.
This value will be checked before the block is allocated
and will be reset after it is freed.
We also check the FBC of $d$ nearby blocks to improve the 
rate of detection.
FBC guarantees that
accessing a freed block is not risk-free.
Its protection rate is analyzed in~\autoref{s:formal}.
In~\autoref{fig:block-remalloc},
we further illustrate how the two kinds of canaries
(FBC and regular heap canary) are set and cleared
when an allocated block is first freed and then allocated again.

Initially in \sys,
we create the memory pool using the \cc{mmap} system call
with the \cc{ANON} flag and
all allocated memories are set to zero
in the Linux environment~\cite{mmapzero}.
We take this advantage and use the zeros as the initial FBC
with the following benefits:

\begin{itemize}[topsep=0pt,itemsep=0pt,partopsep=0pt,parsep=0pt,leftmargin=8pt]
    \item Until being accessed, an unused slot will remain unmapped 
          to the physical memory, decreases memory overhead.
    \item An unused slot is exempted from computing a secure canary value and
          writing to the corresponding memory field.
    \item The whole slot will be checked instead of only the canary bytes,
          increasing the detection rate.
\end{itemize}

\noindent
Treading off the computation cost of the encrypted canaries,
\sys always zeros out the contents of small blocks
and will check the whole block before being allocated to a new object,
bringing both security and computation benefits.

\subsection{Summary and comparison}

\autoref{tab:allocators} summarizes \sys and 
selected state-of-the-art
secure heap allocators
along the two defense lines
that are closely related to \sys
(background discussed
in~\autoref{ss:bg-entropy-alloc}
and~\autoref{ss:bg-uaf-alloc}).

Being an entropy-based allocator,
\sys is inherently closer
to this line of work~\cite{novark2010dieharder,guarder,slimguard}
with a nearly identical set of
heap exploitation protection features
\emph{except} UAF protection.
\sys provides a much stronger security assurance
in the presence of UAF vulnerabilities.
In particular,
\sys addresses the two entropy-loss cases
(discussed in~\autoref{sec:motivation-example} and~\autoref{sec:info-leak})
with RIO (\autoref{sec:rio})
and RBL (\autoref{sec:random-bag-layout}),
respectively,
and hence,
providing much higher effectiveness on UAF mitigation.
In addition,
\sys is designed to actively monitor the integrity of the heap
and watch for UAF attempts, including heap spraying practices
that aim to prepare the heap data and layout for UAF exploits.
\sys achieves this through a synergy of
regular heap canaries (\autoref{sec:regular-canary})
and FBC (\autoref{sec:fbc}).

On the other hand,
\autoref{tab:allocators} also shows a sheer contrast
between entropy-based allocators and UAF-mitigating allocators.
Notably, although providing
a theoretically complete mitigation guarantee toward UAF,
UAF-mitigating allocators~\cite{markus,ffmallo}
significantly impair program efficiency and
are hard to be deployed in time-sensitive use cases.
In contrast,
as will be presented in~\autoref{s:eval},
\sys incurs a significantly lower overhead
that is typical for entropy-based allocators,
making \sys practical and deployable on production systems
if the end-user can tolerate a marginal chance
of protection failure
(less than 10\% in the default setting of \sys,
discussed in~\autoref{s:formal}).

\section{Security Evaluation}
\label{s:formal}

\begin{figure*}[t]
    \centering
    \begin{subfigure}{.32\textwidth}
        % \advance\leftskip-.05\textwidth
        \includegraphics[width=1.0\textwidth]{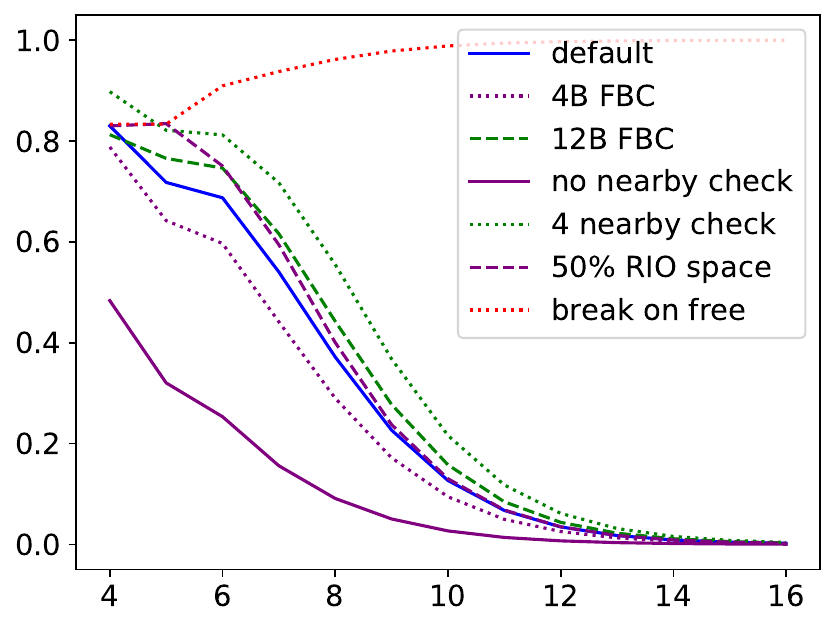}
        \caption{S1 protection rate }
        \label{fig:defense-rate-s1}
    \end{subfigure}
    \begin{subfigure}{.00\textwidth}
    \end{subfigure}
    \begin{subfigure}{.32\textwidth}
    
        % \advance\leftskip-.05\textwidth
        \includegraphics[width=1.0\textwidth]{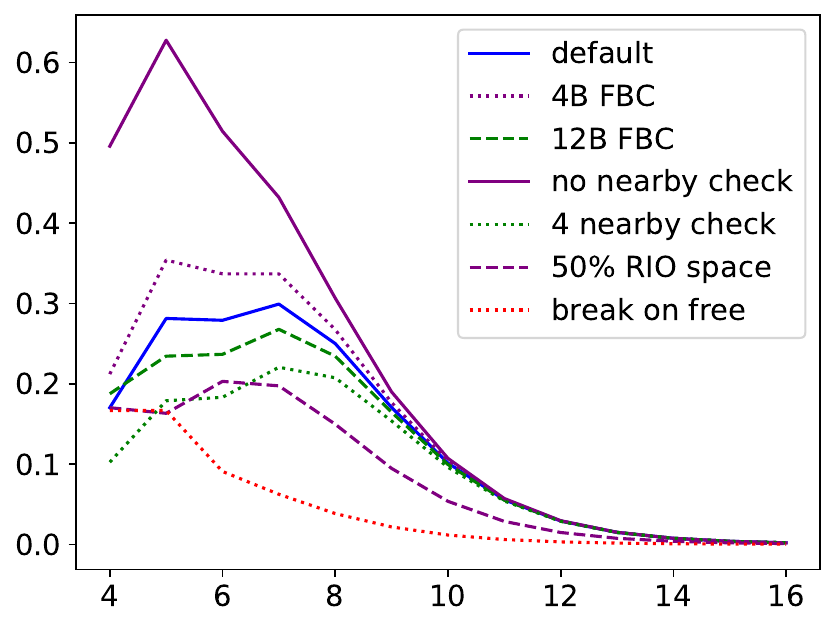}
        \captionsetup{margin={0.0\textwidth,-0.0\textwidth}}
        \caption{S1 attack success rate}
        \label{fig:attack-rate-s1}
    \end{subfigure}
    % group 2
    \begin{subfigure}{.32\textwidth}
        % \advance\leftskip.05\textwidth
        \includegraphics[width=1.0\textwidth]{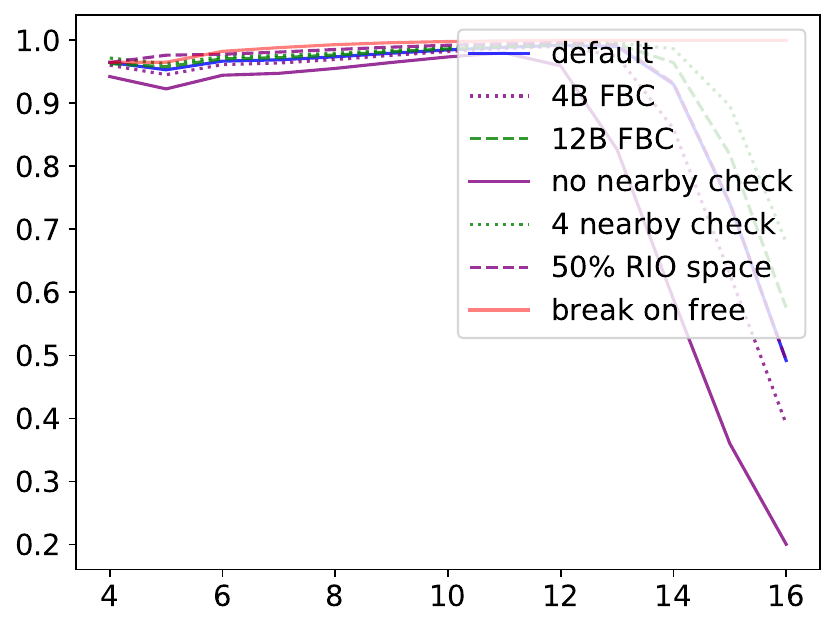}
        \captionsetup{margin={0.0\textwidth,-0.0\textwidth}}
        \caption{S2 protection rate}
        \label{fig:defense-rate-s2}
    \end{subfigure}
    %%%%%%%%%%%%%%%%%%%%%%%%%%%%%%%%%%%%%%%%%%%%%%%%%%%%%%%%%%%%%%%%%%
    \begin{subfigure}{.00\textwidth}
    \end{subfigure}
    %%%
    \begin{subfigure}{.32\textwidth}
        % \advance\leftskip-.12\textwidth
        \includegraphics[width=1.0\textwidth]{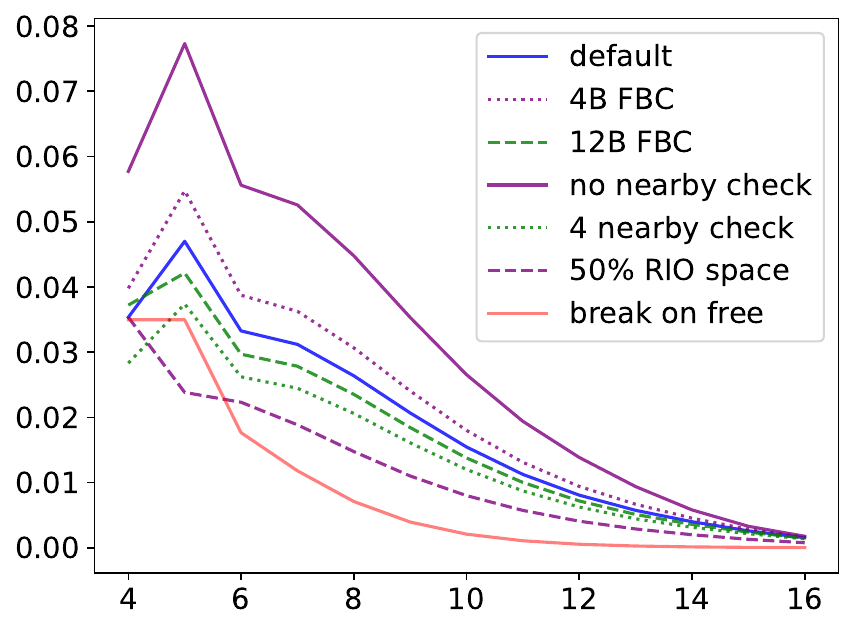}
        \caption{S2 attack success rate}
        \label{fig:attack-rate-s2}
    \end{subfigure}
    % group 3
        \begin{subfigure}{.32\textwidth}
        % \advance\leftskip-.07\textwidth
        \includegraphics[width=1.0\textwidth]{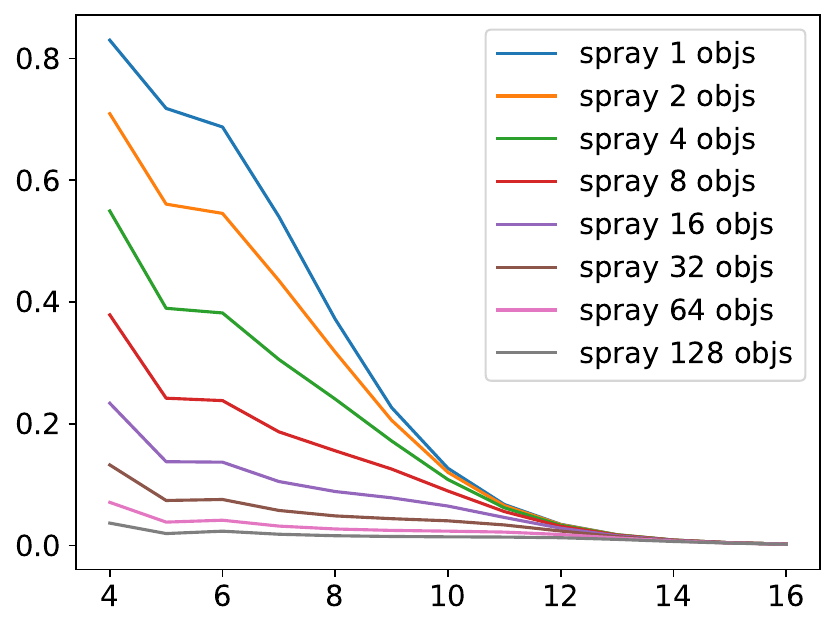}
        \captionsetup{margin={0.0\textwidth,-0.0\textwidth}}
        \caption{S1-spray protection rate}
        \label{fig:spray-rate-s1}
    \end{subfigure}
    \begin{subfigure}{.00\textwidth}
    \end{subfigure}
    \begin{subfigure}{.32\textwidth}
        \advance\leftskip.00\textwidth
        \includegraphics[width=1.0\textwidth]{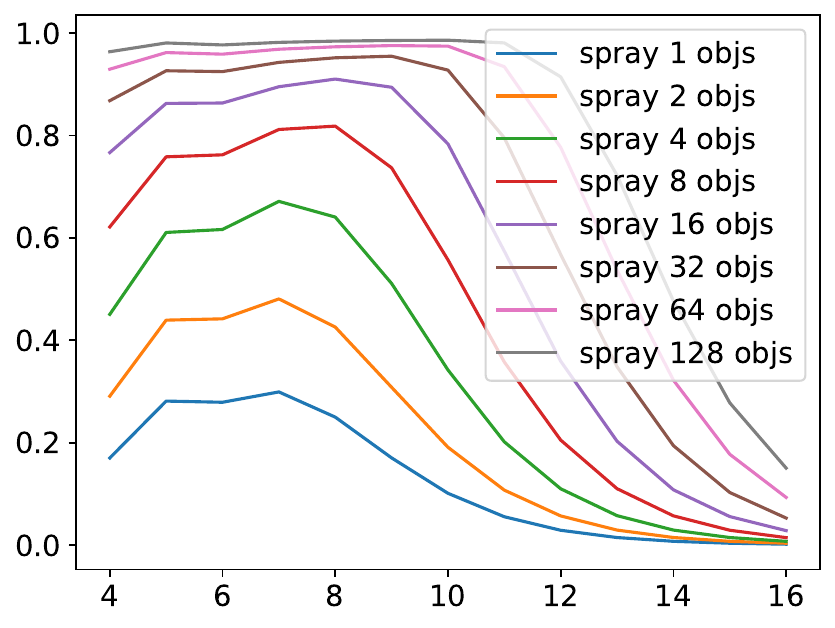}
        \captionsetup{margin={0.0\textwidth,-0.0\textwidth}}
        \caption{S1-spray attack success rate}
        \label{fig:spray-attack-rate-s1}
    \end{subfigure}
    
    %%%%%%%%%%%%%%%%%%%%%%%%%%%%%%%%%%%%%%%%%%%%%%%%%%%%%%%%%%%%%%%%%%
    % group 4
        \begin{subfigure}{.32\textwidth}
        % \advance\leftskip-.1\textwidth
        \includegraphics[width=1.0\textwidth]{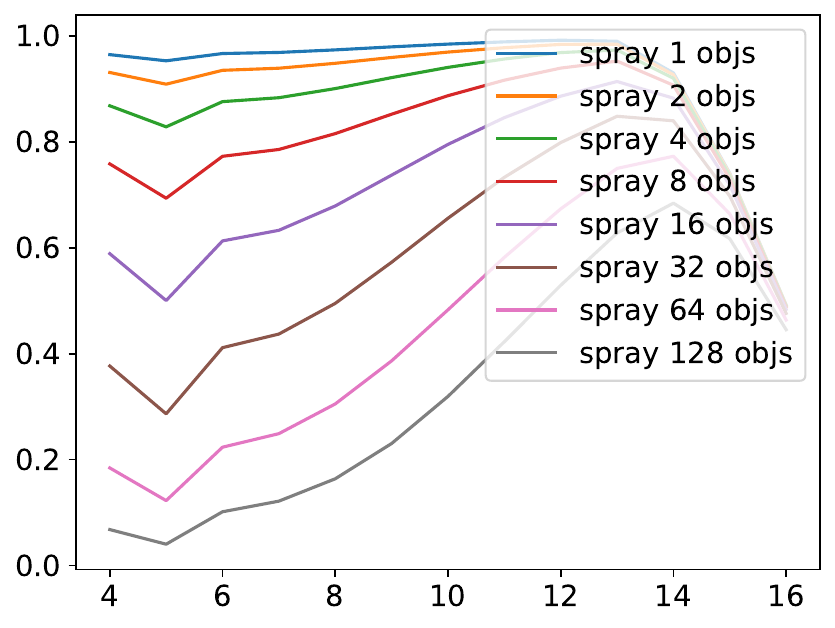}
        
        \caption{S2-spray protection rate}
        \label{fig:spray-rate-s2}
    \end{subfigure}
    \begin{subfigure}{.00\textwidth}
    \end{subfigure}
    \begin{subfigure}{.32\textwidth}
        % \advance\leftskip.1\textwidth
        \includegraphics[width=1.0\textwidth]{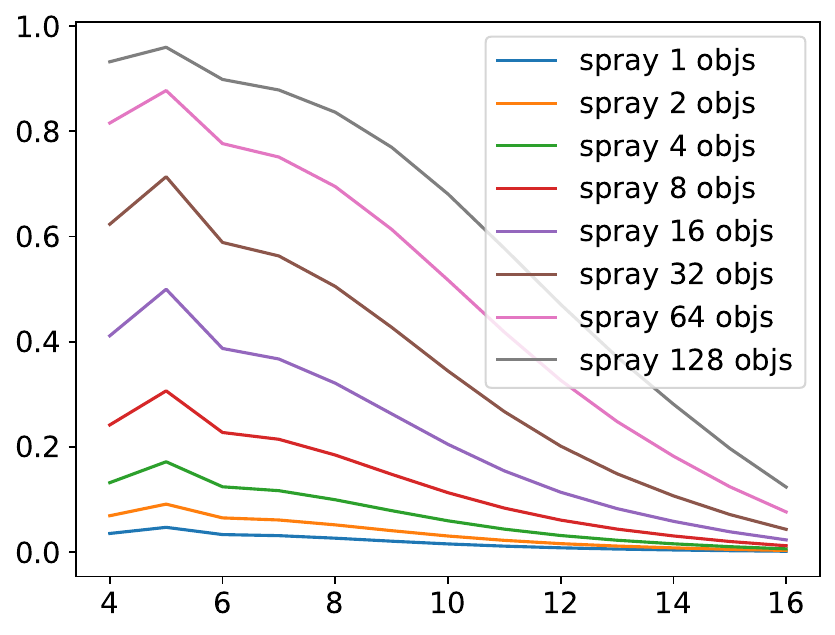}
        \captionsetup{margin={0.0\textwidth,-0.0\textwidth}}
        \caption{S2-spray attack success rate}
        \label{fig:spray-attack-rate-s2}
    \end{subfigure}

    \caption{Parameterized security evaluation (x-axis: logarithmic block size/byte, y-axis: protection/attack success rate/\%).}
\end{figure*}

In this section,
we show the robustness of \sys towards UAF exploitations
in different scenarios.
We first present the results from our formal modeling
and then show how \sys mitigates real-world UAF attacks.

\subsection{Formal analysis}

To mathematically model how \sys provides defense against UAF,
we make the following assumptions for the attacker and target program
(which are consistent with our adversary model in~\autoref{ss:design-adv-model}):
\begin{itemize}[noitemsep,topsep=0pt,leftmargin=20pt]
    \item[\WC{1}] The goal of the attacker is to
        modify the \emph{victim field}, a sensitive field
        (e.g., a function pointer or an \cc{is_admin} flag)
        in a specific type of object,
        a.k.a., \emph{a victim object},
        via memory writes over a dangling pointer
        (i.e., UAF-writes).

    \item[\WC{2}] The attacker can obtain
        a dangling pointer
        through a bug in the program
        at any point of time during execution.

    \item[\WC{3}] The program
        repetitively allocates and frees
        the type of objects targeted by the attacker
        (i.e., victim objects)
        during its execution.
        However, we do not assume that
        each victim object is freed
        before the next victim is allocated.

    \item[\WC{4}] The attacker can
        either indirectly monitor or directly control
        the allocations of victim objects,
        i.e., the attacker knows when a victim object is allocated,
        but does not know the address of the allocation.

    \item[\WC{5}] Any memory writes
        through the dangling pointer is conducted
        after the victim object is allocated.

    \item[\WC{6}] If the intended sensitive field of a victim object
        is overridden, the attack succeeds;
        otherwise, the program continues to execute,
        allowing the attacker to repeat the exploitation effort
        unless detected by \sys (condition \WC{7}).

    \item[\WC{7}] \sys checks FBCs on each heap allocation and
        detects the attack if any FBC is modified.
\end{itemize}

\noindent
To simplify the illustration,
we assume that the above execution logic
is the only code logic that
involves heap management.
In real-world settings,
attackers usually have an even lower success rate as
memory slots can be allocated to other objects,
which gives \sys more chances to check FBCs and
detect UAF attempts.

In addition,
in reality, there will be objects
that are not of the victim type but
bear similar sizes and hence,
will also compete for the same block
with victim objects.
Such objects are called \emph{irrelevant objects} and
their prevalence and longevity at run time affects a realistic attacker's strategy in
attacking \sys.
Below, we introduce two simple and yet realistic strategies
a reasonable attacker might consider:

\PP{S1: Repeat UAF-writes
through the same dangling pointer}
If an attacker is confident that
the memory block pointed to by the dangling pointer
is not taken by a long-living irrelevant object,
the attacker may
prefer to keep re-using that dangling pointer
for future attack attempts and
hope that a victim object is allocated
in that block.

\PP{S2: Repeat UAF-writes with freshly obtained dangling pointers}
If an attacker worries that irrelevant objects
can be long-lasting
(i.e., holding a block that may be allocated
to victim objects),
the attacker may prefer to use a fresh dangling pointer per each attempt.
However, we note that this is only an attack strategy and we assume that there is no heap allocation other than victim objects.

For each of the scenarios above,
we discuss the rates of whether the attacker adopts the heap spray technique or not separately.
We introduce a total of four scenarios.
We refer readers to~\autoref{s:app-formal} for a detailed explanation of them.

\subsection{Parameterized protection rates}

In this section,
we illustrate how the protection and attack success rates
vary with different parameter configurations
using the two attack strategies.
Assuming the victim field is a pointer of 8 bytes,
The set of tunable parameters include
\WC{1} block size,
\WC{2} FBC length,
\WC{3} RIO entropy,
\WC{4} break on free, and
\WC{5} number of free blocks with FBCs checked.

For FBC-checking, we take 0 
and 4 nearby blocks as a comparison to the default setting (2 nearby blocks).
For FBC length, we take 4 and 12 bytes as a comparison with the default (8 bytes).
For random offset entropy, 
we reserve 50\% of block size for RIO compared to the default (25\%).
We also evaluate the influence
on the two rates with 
break-on-free.

We provide estimates of
the rates by assuming that
an attacker re-tries an attack
for 500 times at max
even if the attack is still
not detected by the defender.
We also assume that overwriting the heap canary does not trigger any alarm.
We analyze the protection and attack success
rates of blocks with sizes ranging from
16 bytes to 64K bytes
as larger blocks are resistant to UAF attacks.
All these simplifications favor the attacker.

\autoref{fig:defense-rate-s1} shows how
the protection rate
changes with a different set of tunable parameters
while
\autoref{fig:attack-rate-s1} shows how
the attack success rate changes using
strategy S1.
We observe that adopting a more secure setting increases 
the protection rate and decreases the attack success rates for both small and large blocks.
However, as the block size increases, 
both rates decrease as it is less likely to overwrite either the target field or FBC.
Both the attacker and the defender are likely not to succeed, and the rates are thus not summed up to one
(in fact, not even close to one).

\autoref{fig:defense-rate-s2} shows how
the protection rate
changes with the same set of tunable parameters
while
\autoref{fig:attack-rate-s2} shows how
the attack success rate changes using
strategy S2.
Similarly, we observe that a more secure setting increases 
the protection rate and decreases the attack success rates.
However, the attacker would leave numbers of overwritten FBC 
in the memory pool in this strategy making the protection rate 
higher.

Note that other entropy-based
allocators provide zero
protection
as they do not detect failed
UAF-write attempts.

\subsection{Protection rates with heap spray}
Heap spraying boosts attackers'
chance of overwriting a sensitive field
in a victim object and hence,
increases the attack success rate and
decreases the protection rate.
(see detailed strategies
and formal analysis in~\autoref{ss:strategy-3} and~\autoref{ss:strategy-4}).
We use the default settings of \sys and
evaluate how both attack and defense rates change of strategies (S1 and S2)
with different numbers of sprayed objects and object sizes.
\autoref{fig:spray-rate-s1} and~\autoref{fig:spray-attack-rate-s1}
show the protection and attack success rates of reusing the same pointer (strategy S1-spray).
\autoref{fig:spray-rate-s2} and~\autoref{fig:spray-attack-rate-s2} show the protection and attack success rates using a fresh pointer (strategy S2-spray).

Spraying the heap with target variables diminishes the random allocation
entropy,
thus decreases the protection rate and increases the attack success rate 
as FBC will not be overwritten if the pointed block is allocated thus making the attack 
not detectable. 
RIO entropy is not influenced by heap spray.
An effective protection can be achieved by adopting a more secure configuration
(e.g., checking more nearby FBCs,
larger RIO or block entropy),
with 
marginal performance degradation (see~\autoref{ss:result-parameters})
to make the spraying less effective.

\subsection{Illustrate the protection rates}
We take the mRuby issue 4001 (shown in~\autoref{code:uaf-example}) as an example and show how its protection rate is computed. 
The size of object \cc{mrb_io} is 16 bytes,
In each run, the attacker's goal is to overwrite 4 bytes of it.
With the default settings ($r = 256$, $s = 32$ to store a 16-byte object), the attack success rate of each trial is approximately $0.002$, and the probability of overwriting FBC in a free block is approximately $0.16$.
In~\autoref{tab:protection-exp}, we show how the rates change as the number of attack rounds goes up.
The attack is 64\% likely to be detected if the attacker adopts the first attack strategy and 95\% likely to be detected using the second strategy after running it 500 times. 

\begin{table*}[t]
    \centering
    \small
    \begin{tabular}{c|cccccc|cccccc}
        \toprule
        & \multicolumn{6}{c}{Attack Strategy 1} & \multicolumn{6}{c}{Attack Strategy 2} \\ \midrule
         round & 1 & 5 & 10 & 50 & 100 & 500 & 1 & 5 & 10 & 50 & 100 & 500 \\
         \midrule 
         $p_{protection}$ & 1.4\% & 4.1\% & 7.4\% & 28\% & 43\% & 64\% & 0.8\% & 12\% & 37\% & 95\% & 95\% & 95\% \\
         $p_{attack}$ & 1.2\% & 2.6\% & 4.4\% & 15\% & 24\% & 35\% & 1.2\% & 2.6\% & 4.0\% & 5.5\% & 5.5\% & 5.5\% \\
    \bottomrule
    \end{tabular}
    \caption{Protection and attack success rates of attack rounds in mRuby issue 4001 using the two strategies.}
    \label{tab:protection-exp}
\end{table*}

\subsection{Defending against real-world CVEs}
\label{sec:real-world-example}

In this section, we take seven recent vulnerabilities and compare how \sys, Guarder,
DieHarder and SlimGuard fare in defending against them.
We select vulnerabilities based on the following criteria:
\begin{itemize}[topsep=0pt,itemsep=0pt,partopsep=0pt,parsep=0pt,leftmargin=8pt]
    \item On the Linux platform and can be mitigated in the user space (i.e., not a Linux kernel bug);
    \item Can be deterministically triggered (i.e., not racy);
    \item Public exploit is available and the exploit breaks the program information integrity (i.e., not only causing DoS).
\end{itemize}

\begin{figure*}[h]
    \hfill\begin{minipage}[c]{.7\linewidth}
    \input{code/cve-zval.c}
    \end{minipage}
    \caption{Type definition of \protect\cc{zval}.}
    \label{fig:cve-zval}
\end{figure*}

\begin{table*}[t]
    \centering
    \small
    \begin{tabular}{llcccc|cc}
        \toprule
        \textbf{Vulnerability} & \textbf{Attack pattern} & \textbf{\cite{guarder}} & \textbf{\cite{novark2010dieharder}} & \textbf{\cite{slimguard}} & \textbf{\sys} & \textbf{\cite{typeaftertype}} & \textbf{\cite{cling}} \\
        \midrule 
        CVE-2015-6831 & DP $\rightarrow$ Write & \LEFTcircle & \Circle & \Circle & \CIRCLE & \Circle & \Circle \\
        CVE-2015-6834 & DP $\rightarrow$ Write & \LEFTcircle & \Circle & \Circle & \CIRCLE & \Circle & \Circle \\
        CVE-2015-6835 & DP $\rightarrow$ Write & \LEFTcircle & \Circle & \Circle & \CIRCLE & \Circle & \Circle \\
        CVE-2015-6835 & DP $\rightarrow$ Write $\rightarrow$ sleep  & \Circle & \Circle & \Circle & \CIRCLE & \Circle & \Circle \\
        CVE-2020-24346\xspace\xspace\xspace & DP $\rightarrow$ Write & \Circle & \Circle & \Circle & \CIRCLE & \Circle & \Circle \\
        Python-91153 & DP $\rightarrow$ Write & \Circle & \Circle & \Circle & \CIRCLE  & $\blacksquare$ & \Circle \\        
        mruby-4001 & DP $\rightarrow$ Write & \Circle & \Circle & \Circle & \CIRCLE & $\blacksquare$ & $\blacksquare$ \\
        CVE-2022-22620 & DP $\rightarrow$ Read & \Circle & \RIGHTcircle & \Circle & \circledbullet & \Circle & \Circle \\
    \bottomrule
    \end{tabular}
    \caption{Sumsmary of how different memory allocators
    defend against eight exploitation techniques on seven vulnerabilities. Vanilla BIBOP allocator and Scudo~\cite{scudo} are vulnerable to all attacks and behave similarly to Guarder~\cite{guarder} (DP: dangling pointer, \Circle: no defense, \LEFTcircle: detect at the end of execution, \RIGHTcircle: defense via zero-out, \CIRCLE: detect via FBC change, \protect\circledbullet: non-deterministic leak (RIO), $\blacksquare$: thwart the exploitation ability).
    }
    \label{tab:cves}
\end{table*}

Out of the seven vulnerabilities,
six are UAF-write bugs and
one (CVE-2022-22620) is a UAF-read only bug.
We also found seven exploits against these vulnerabilities
(two exploits for CVE-2015-6835 with different attack patterns).
All CVEs except Python-91153 and mruby-4001
can cause arbitrary code execution (ACE)
if properly exploited.
However,
a powerful attack (e.g., ACE by overriding a function pointer)
can succeed when the target object is precisely allocated
to a freed memory chunk that is still referred to by a dangling pointer\footnote{An
attacker may attack blindly,
e.g., overriding a code pointer through the dangling pointer
regardless of whether it points to a target object or not.
This will have three consequences:
1) corrupting FBC,
which may cause the attack to be detected upon future \cc{malloc}s,
2) overwriting a wrong field due to RIO
which may cause the program to enter a weird state (e.g., crash),
3) a successful attack.
If the program can be recovered from a weird state automatically
(e.g., crash resilience),
the attacker can retry the same attack and eventually case 1 or case 3 will occur. 
However,
without the probabilistic detection on UAF attempts by \sys,
only case 3 will occur.}.
\sys can mitigate attacks by reducing the chance that
a target object is referred to by a dangling pointer.
Evaluation results of the eight exploits are in~\autoref{tab:cves}.

\PP{Entropy-based allocators}
\sys can thwart all UAF-write attacks evaluated,
Guarder can detect three exploits
by recognizing double-free attempts, but
DieHarder and SlimGuard fail to thwart these exploits.
For the UAF-read attack,
\sys uses RIO to stop the attacker from
reusing the memory chunk with accurate object alignment,
causing the data read by the dangling pointer to be not sensible.
DieHarder zeros out the memory after it is freed, which is effective if the attacker tries to over-read a freed block.

\PP{Context-based allocators}
While we expect context-based allocators to demonstrate
strong and stable protection,
some of the exploits, unfortunately, hit on
certain blind spots in Cling and TypeAfterType
by accident.

In the case of Cling,
if both the dangling pointer and the target object
(i.e., the object an attacker hope to corrupt)
are allocated through the same
multi-layer function call sequence,
they are considered to fall under the same allocation context,
causing the target object to be possibly accessed by the dangling pointer.
We will illustrate this limitation
through the CVE-2015-6835 case study
presented later.
In fact,
all examined CVEs, except mruby-4001,
hit this limitation of Cling.
Cling mitigates mruby-4001
by limiting the attacker to target
objects of type \cc{mrb\_io},
which prevents the attacker from
creating a powerful attack primitive.

TypeAfterType can unpack \cc{malloc} wrappers
with an arbitrary number of layers until
it finds a \cc{sizeof(T)} in the function argument,
and an ID \cc{i} is given to each allocation site of \cc{T}.
The tuple (\cc{i},\cc{T}) makes the allocation context,
and all memory allocations through this call sequence
will be allocated from a memory pool dedicated to this context.
However, 
if the dangling pointer and the target object
share the same context in an exploit,
UAF can still occur.
We will illustrate this limitation
through the CVE-2015-6835 case study.
TypeAfterType mitigates
Python-91153 by limiting the
target object to be a reallocated \cc{string},
and mruby-4001 by limiting the
target object to be an \cc{mrb\_io}.

\PP{Case study: CVE-2015-6835}
This CVE is a UAF bug in the PHP session deserializer,
which reconstructs a session from a serialized string.
(\autoref{fig:cve-poc}).
An attacker can exploit this vulnerability 
to control a dangling pointer to a freed \cc{zval} object.
This is possible as the return value (a \cc{zval} pointer)
\cc{php_var_unserialize}
is freed in its caller
without noticing that the same pointer might also be stored
in a global variable \cc{SESSION_VARS}.

The \cc{zval} type, unfortunately,
is a reference-counting wrapper
over nearly all other objects
involved in the PHP engine
(see definition of \cc{zval} in~\autoref{fig:cve-zval}).
Therefore,
an attacker might
corrupt any \cc{zval} object
that may be reallocated to the freed slot.
They can simply uses the
\cc{echo(..)} function to dump
a newly allocated \cc{zval} in the freed memory.

\emph{1) Protection from entropy-based allocators.}
\sys checks FBC on every \cc{malloc()}.
In this exploit,
when the attacker tries to use the dangling pointer
in \cc{zend_echo_handler},
its \cc{refcount} field is increased,
causing the FBC to be modified.
This enables \sys to detect the UAF attempt
when the corrupted slot or a nearby slot
is about to be reallocated.
If the \cc{refcount} change does not corrupt FBC
(simulated by disabling the FBC check)
and this corrupted block is reallocated,
\sys can still stop the exploit as
RIO causes misalignment between the dangling pointer and new object,
causing the \cc{type} field to have value \cc{UNKNOWN}
and prevents echo printing.

Guarder and DieHarder try to mitigate this attack by random allocation: 
hoping the new object will not be referred to by a dangling pointer.
However, our experiment shows that Guarder fails
if the attacker re-runs the attack multiple times 
or spray the heap with victim objects.
SlimGuard fails to provide protections
as it always allocates the most recently
freed objects to the program.
It does not implement the claimed random allocation feature and
does not have any other security features that could detect UAF.
DieHarder zeros out the memory chunk that
stops information leakage of the freed \cc{zval},
but it cannot prevent an attacker to corrupt
the newly allocated \cc{zval} over the freed chunk.

\emph{2) Protection from context-based allocators.}
In this exploit, 
both the dangling pointer and the target object
(i.e., the object the attacker wish to dump information
via \cc{zend_echo_handler})
are allocated by the
the same multi-layer \cc{malloc} wrapper:
\cc{php_var_unserialize}$\rightarrow$\cc{emalloc}$\rightarrow$\cc{malloc}.
This is critical to understand
why Cling and TypeAfterType fail to mitigate this exploit.

For Cling,
this \cc{malloc} wrapper implies that
the allocation of many \cc{zval} objects will be sharing
the same context
(measured by the two innermost return addresses on the call stack).
This leaves the dangling pointer plenty of
candidate objects to refer to
after several rounds of deserialization in PHP.
TypeAfterType can inline \cc{malloc} wrappers
but the inlining stops at \cc{php_var_unserialize}
because it sees the \cc{sizeof(zval)} argument in \cc{emalloc}
and hence,
will allocate all \cc{zval} objects
originating from this \cc{malloc} wrapper from the same pool.
Unfortunately, the dangling pointer is also allocated this way,
enabling UAF among the dangling pointer to other \cc{zval} objects as well.

\PP{Summary}
The combination of random allocation and delayed free-list provided
by previous entropy-based allocators focus on one-time attacks only,
Hence, repeating the same attack
is a simple yet effective solution to undermine their protection.
Context-based allocators, on the other hand,
might fail to detect UAF among objects allocated of the same context.
While these results highlight the effectiveness of \sys, 
we note that information leakage through 
corrupted pointers might diminish this guarantee,
which is discussed in~\autoref{s:discuss}. 

\vspace{-5pt}
\section{Performance Evaluation}
\label{s:eval}

In order to evaluate the performance and memory
overhead of these allocators,
we run various benchmarks 
trying to provide a complete understanding of their performance.
We firstly run two macro benchmarks -- PARSEC and SPEC (\autoref{ss:macro-benchmarks}),
and then use the mimalloc-bench and glibc micro-benchmark to evaluate the performance of running two 
most frequent heap memory management functions: \cc{free()} and \cc{malloc()} 
(\autoref{ss:micro-benchmarks}).
We then evaluate their performance on real-world programs using two servers: Nginx and Lighttpd, and two databases: Redis and SQLite (\autoref{ss:server-benchmarks}).
We then discuss how multi-threading impairs each 
of them performance (\autoref{ss:thread-benchmarks}).
In the end, we show how different parameter values 
influence the performance of \sys (\autoref{ss:result-parameters}).

\PP{Experiment setup}
Experiments are performed on both x86 and AARCH servers for macro benchmarks.
The performance of benchmarks is measured only on the x86 server.
The x86 server is configured with 64-bit 160-core 2.40GHz Intel Xeon E7-8870 CPUs (x86 architecture)
with 1TB system memory.
We set up the AARCH server on Amazon Web Service (AWS),
using the \cc{im4gn.4xlarge} machine
with 16 vCPU cores and 64 GB memory.
On both machines, benchmarks are measured in the Docker environment with 
Debian 11, kernel version 5.15.0.
We measure the overheads using the GNU \cc{time} binary~\cite{linux-time} and
setting the \cc{LD\_PRELOAD} environment variable to substitute the system default allocator.

We obtain SlimGuard, Guarder, and DieHarder from their corresponding GitHub repository.
We use SlimGuard with commit \cc{81f1b0f} as
a later erroneous commit prevents us from using \cc{LD\_PRELOAD} 
to replace the system allocator.
We use the up-to-date version of the other two memory allocators (Guarder: \cc{9e85978}, DieHarder: \cc{640949f}).
In order to provide a fair result,
we reduce the allocation entropy bit of
Guarder to eight (same as the default value of SlimGuard and \sys).
We also disable DieHarder from zeroing out freed blocks
(this actually slightly accelerates DieHarder).

\sys is measured with the settings of checking two nearby blocks ($d = 2$),
10\% random guard page, 
and taking 1/4 of the block size as random offset entropy ($e = 0.25b$).
For blocks smaller than a memory page (4096 bytes),
we zero it out and take the whole block as FBC.
For blocks larger than a memory page, 
we set an 8-byte FBC ($c = 8)$ in the corresponding blocks.
We set the heap canary length to be one byte ($\iota=1$) following the design of SlimGuard and Guarder. 
All reported times and memory usage are normalized using the baseline (glibc) output.
We use geometric averages to compute average overheads and
report the means and standard deviations of five runs.

\begin{table*}[t]
    \centering
    \small
    \begin{tabular}{c|cccc|cccc}
        \toprule
                  & \multicolumn{4}{c}{Run-time Overhead} & \multicolumn{4}{c}{Memory Overhead}                                                          \\
                  \midrule
                  & \multicolumn{2}{c}{\textbf{x86}}
                  & \multicolumn{2}{c}{\textbf{AARCH}}
                  & \multicolumn{2}{c}{\textbf{x86}}
                  & \multicolumn{2}{c}{\textbf{AARCH}}
                  \\
                  \cmidrule{2-9}
                  & SPEC & PARSEC & SPEC  & PARSEC 
                  & SPEC & PARSEC & SPEC  & PARSEC \\\midrule
        \textbf{\sys}  & \textbf{12\% } & \textbf{2.8\%} & \textbf{16\% } & \textbf{1.8\%}   & \textbf{37\% } & \textbf{27\% } & \textbf{38\%}  & \textbf{28\% }  \\
        SlimGuard & 17\%  & 4.4\%  & 7.7\% & 2.6\%    & 57\%  & 23\%  & 57\%  & 24\%   \\
        DieHarder & 31\%  & 2.1\%  & -            & 2.5\%    & 59\%  & 21\%  & -          & 21\% \\
        Guarder   & 3.5\%   & 2.4\%  & -            & -             & 56\%  & 58\% & -          & -       \\\bottomrule
    \end{tabular}
    \caption{Normalized overheads
    on SPEC and PARSEC benchmarks.
    }
    \label{tab:results-summary}
\end{table*}

\subsection{Macro benchmarks}
\label{ss:macro-benchmarks}

\PP{PARSEC}
We first evaluate the performance of \sys using the PARSEC~\cite{parsec} benchmark.
We exclude three network tests (netdedup, netferret, and netstreamcluster)
and one test (x264) that fails to compile in the baseline scenario,
and only report the result of the rest 12 benchmarks.
Additionally, we exclude ``raytrace'' from execution for the AARCH sever
as it cannot compile.
We refer to each PARSEC test using the first three letters of its name.

\PP{SPEC CPU2017}
We use SPEC CPU2017~\cite{spec} version 1.1.9.
We report the results of 12 C/C++ only tests
in both ``Integer'' and ``Floating Point'' test suites with the
default OpenMP settings of four parallel threads.
All reported SPEC overheads are ``reportable'' 
following the SPEC documentation~\cite{spec}.

\PP{Results}
We measure the performance of \sys and three other entropy-based 
allocators (SlimGuard, Dieharder and Guarder) on
both benchmarks with the x86 machine.
On the AARCH machine, we exclude Guarder from analysis,
noticing that Guarder relies on AES-NI~\cite{aesni},
an Intel CPU extension, not supported on AARCH machines.
We report the average and standard deviations of 
overheads in~\autoref{tab:results-summary}.
Averages and standard deviations of each test in 
both settings are shown in~\autoref{ss:overheads}.
Missing columns in the figures indicate the corresponding execution runs into error.
A complete list of erroneous 
executions and explanations is listed in~\autoref{s:failed-tests}.

For the SPEC benchmark,
on the x86 machine, \sys introduces 11.5\% run-time overhead, 
smaller than two allocators -- SlimGuard and DieHarder,
and introduces the least memory overhead at 37.4\%.
On the AARCH machine, \sys introduces similar 15.5\% 
run-time overhead,
larger than the SlimGuard due to the fact that SlimGuard fails to 
run tests with frequent heap memory management operations.
Running the PARSEC benchmark gives similar results,
with smaller memory and run-time overheads.

We observe \sys and other memory allocators introduce 
larger overheads for tests with frequent heap memory management operations,
for example, ``ded'' in PARSEC and ``620'' in SPEC.
We investigate the costs of running \cc{malloc()} and \cc{free()} in the following
section.

\begin{table}
\parbox{.47\linewidth}{
\centering
\small
\begin{tabular}{cccc}
        \toprule
        & \begin{tabular}[c]{@{}c@{}}Run-time \\ Overhead\end{tabular} & \begin{tabular}[c]{@{}c@{}}Memory \\ Overhead\end{tabular} \\
        \midrule
        \sys & 189\% & 343\% \\
        SlimGuard & 298\% & 250\% \\
        DieHarder & 229\% & 92\% \\
        Guarder & 56\% & 980\% \\
        \bottomrule
        \end{tabular}
        \caption{Normalized run-time and memory overheads of running mimalloc-benchmark}
        \label{tab:mimalloc}
}
\hfill
\parbox{.47\linewidth}{
\centering
\small
\begin{tabular}{lcc}
        \toprule
        & \begin{tabular}[c]{@{}c@{}}$\Delta$ Memory \\ Overhead\end{tabular} & \begin{tabular}[c]{@{}c@{}}$\Delta$ Run-time \\ Overhead\end{tabular} \\
         \midrule 
         0 Nearby & -0.05\% & -0.42\% \\
         4 Nearby & -0.13\% & +0.4\% \\
         4B Random N & -12.35\% & -0.33\% \\
         12B Random N & +68.23\% & +0.73\% \\
         50\% Entropy & +9.57\% & +0.49\% \\
        \bottomrule
    \end{tabular}
    \caption{
        Normalized memory and run-time Overhead changes compared with the default Settings.
    }
    \label{tab:result-parameters}
}
\end{table}

\subsection{Micro benchmarks}
\label{ss:micro-benchmarks}

\begin{figure}[!htb]
\minipage{0.48\textwidth}
  \centering
    \includegraphics[width=\linewidth]{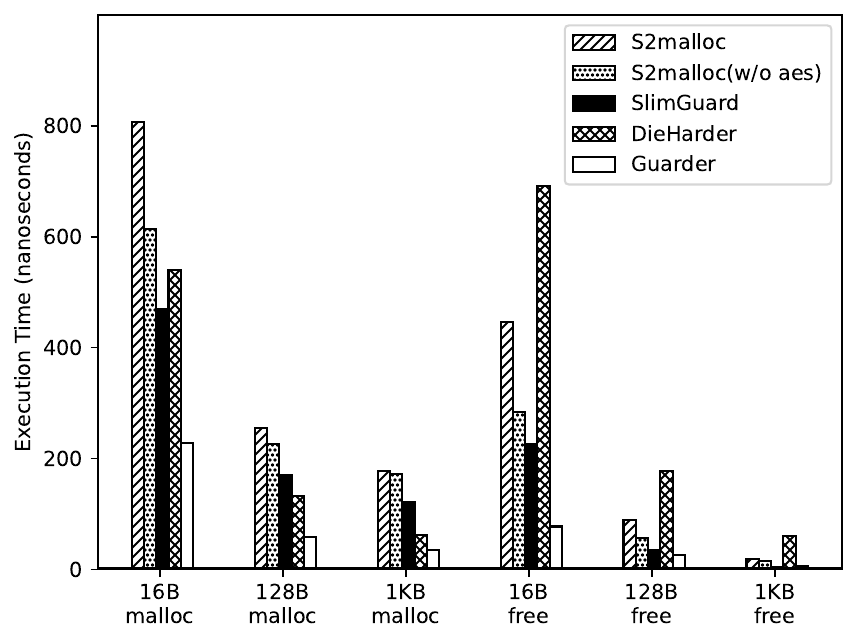}
    \caption{Execution time of glibc-simple
    }
    \label{fig:micro-result}
\endminipage\hfill
\minipage{0.48\textwidth}%
      \centering
    \includegraphics[width=\linewidth]{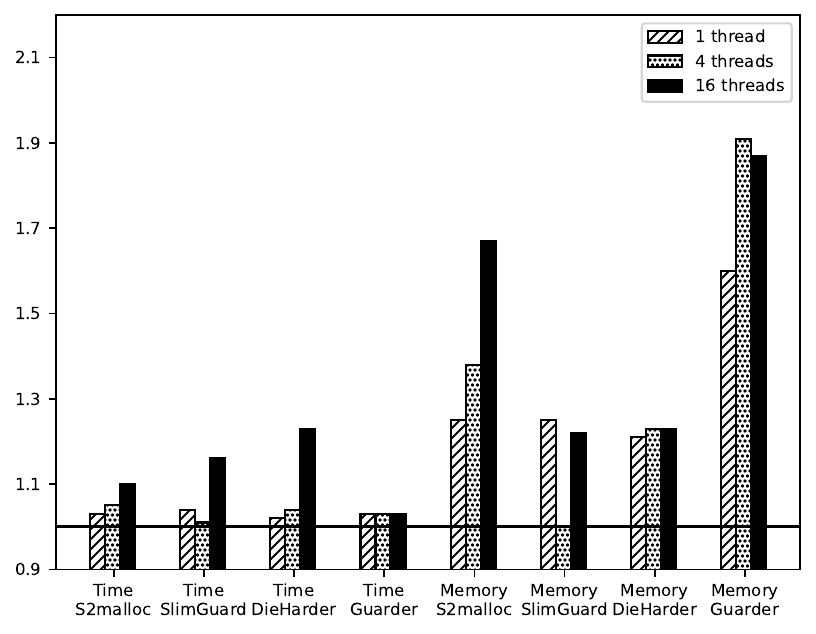}
    \caption{Run-time and memory overheads of running PARSEC with multi-threads
    }
    \label{fig:thread}
\endminipage
\end{figure}

To further understand the overheads introduced by \sys,
we investigate its performance using mimalloc-bench~\cite{mimalloc-bench}, composed of real-world and calibrated programs that allocate and free heap memory frequently. 
The results are shown in~\autoref{tab:mimalloc}.
Individual results are reported in~\autoref{ss:mimalloc-app}.
Running three tests with SlimGuard never returns (marked as "\cc{-}") in~\autoref{ss:mimalloc-app}. 
They are excluded from computing the SlimGuard average overheads.

All secure memory allocators incur larger overheads compared to running real-world (see~\autoref{ss:server-benchmarks}) or general-purpose benchmarks as mimalloc-bench tests operate the heap memory in a biased frequent way and some tests (e.g., "leanN" generates the largest run-time overhead with \sys) counts the CPU ticks instead of seconds of finishing each call. 

We take one test, \cc{glibc-simple} from mimalloc-bench, 
 from the glibc micro-benchmark suite~\cite{glibc-bench} 
to further investigate the delays incurred in the two most common
heap object management functions -- \cc{malloc()} and \cc{free()}.
The test times the execution of 
allocating and freeing a large number of 
blocks of a given size.
We modify the benchmark and 
monitor the execution time of calling \cc{malloc()} and \cc{free()} separately.
To investigate the time consumption of different sizes,
we vary the block size \cc{S} to be 16B, 128B, and 1KB,
and change the number of allocated blocks \cc{N} correspondingly,
so that the total allocation size (\cc{N * S}) is always 1000 MB.
Results are presented in~\autoref{fig:micro-result},
and is the average of 100 runs.

Generally, \sys takes more time to execute \cc{malloc()} than all other compared 
memory allocators,
and takes less time to execute \cc{free()} than DieHarder but longer time than 
Guarder and SlimGuard.
However, a significant overhead comes with our cryptographically secure canary implementation,
which should be a standard adopted by all memory allocators.
Although using hardware acceleration, 
the canary value is computed in each \cc{malloc()} and \cc{free()} calls,
introducing a nonnegligible computation tax.
After disabling this feature and using a fixed value as the canary,
following the implementation of Guarder,
although the execution time of both calls is still longer than Guarder and SlimGuard,
it is comparable to others and 
the increased overhead is expected as \sys introduces extra security
guarantees. 
For example, in 16B \cc{malloc} call, \sys is 31\% slower than SlimGuard and is 26\% 
slower than SlimGuard in 16B \cc{free}.

\subsection{Performance on real world programs}
\label{ss:server-benchmarks}

To evaluate the performance of \sys in real-world environments, we run two servers: Nginx (1.18.0), and Lighttpd (1.4.71), and two databases: Redis (7.2.1), and SQLite (3.25.2) on the x86 machine.
We use ApacheBench (ab)~\cite{ab} 2.3 to test the throughput and delays using the Nginx default root page, of 613 bytes,
as the requested page with 500 concurrent requests. 
On Redis, we use the same settings as its performance is measured in mimalloc-bench~\cite{mimalloc-bench}. 
We use sqlite-bench~\cite{sqlite-bench} to measure the performance of SQLite. 
We report the results of performing random read and write operations 
in~\autoref{tab:results-server} and~\autoref{tab:results-database}.
We observe that applying \sys on these programs results in minimal throughput influence (even better throughput on Nginx and Redis). Running \sys delays the request response time on Nginx but not on Lighttpd. Applying all secure memory allocators increases memory consumption.

\begin{table}[t]
\centering
\small
\begin{tabular}{lrrr|rrr}
\toprule
& \multicolumn{3}{c}{Nginx} & \multicolumn{3}{c}{Lighttpd} \\
& Throughput & Memory &  p50   &  Throughput  &  Memory  &  p50    \\\midrule
\sys & 9705.369 & 7867.2 & 56 & 11050.830  & 9425.6  & 44\\
Guarder       & 9496.210            & 13724  & 52 & 11146.580  & 11088.8 & 44\\
SlimGuard     & 6159.014           & 4935.2 & 81   & 11153.358 & 6428.0    & 44\\
DieHarder     & 8769.120            & 7396.8 & 57 & 11128.114 & 13626.4 & 44\\
Glibc         & 9564.754           & 3400.0   & 52   & 10742.708 & 5069.6  & 44\\\bottomrule
\end{tabular}
\caption{Memory consumption (KB), throughput (requests/second), and delays (msec) for servers}
\label{tab:results-server}
\end{table}

\begin{table}[t]
\centering
\small
\begin{tabular}{lrr|rr|rr}
\toprule
& \multicolumn{2}{c}{Redis} & \multicolumn{2}{c}{SQLite-Read}& \multicolumn{2}{c}{SQLite-Write} \\
              & Throughput  & Memory &  Throughput  &  Memory  &  Throughput &  Memory \\\midrule
\sys          & 218460.294          & 44688.8 & 771247.8791 & 17234.4 & 25146.73118 & 208592.0   \\
Guarder       &          221245.016 & 46548.8 & 790263.9482 & 17451.2 & 25735.65366 & 193408.8 \\
SlimGuard     &          219986.106 & 47397.6 & 766518.4731 & 14921.6 & 24456.21604 & 137659.2 \\
DieHarder     &          221733.848 & 52419.2 & 781983.1092 & 19115.2 & 24613.93050 & 377160.8 \\
Glibc         &          218155.764 & 50908.8 & 796812.7490  & 5914.4  & 25439.72566 & 132984.8 \\\bottomrule
\end{tabular}
\caption{Memory Consumption (KB) and throughput (requests/second) on databases}
\label{tab:results-database}
\end{table}

\subsection{Performance with multi-threading}
\label{ss:thread-benchmarks}

We run the memory allocators on the PARSEC benchmark 
with 4 and 16 threads separately using the x86 machine.
We exclude the test ``ray'' from analysis as it cannot be executed with multiple threads
and ``vip`` as running it using Guarder with 16 threads
causes a segmentation fault.
Results are reported in~\autoref{fig:thread}.

We observe that as the number of threads increases,
\sys gradually introduces more run-time and memory overheads,
as we use atomic instructions and maintain per-thread metadata.
SlimGuard and DieHarder use single global metadata,
and use \cc{lock} to achieve multi-thread compatibility.
While increasing the number of threads does not introduce
extra memory overheads on the one hand,
\cc{lock} introduces more run-time overheads on the other hand.
Guarder uses per-thread metadata but fails to use atomic instructions
to update the metadata, 
causing racing conditions if multiple threads are handling 
adjacent blocks.

\subsection{Influence with different parameters}
\label{ss:result-parameters}

In addition to the default settings,
we also measure how different parameters, namely,
nearby checking range, random offset entropy, and RIO entropy,
influence the run-time and memory overheads.
For the nearby checking range, we take 0 and 4 blocks as a comparison to the default setting: 2.
For random allocation entropy, we take 4 bytes and 12 bytes as a comparison to the default 8 bytes.
For random offset entropy, we reserve 50\% of block size for RIO compared to the default 25\%.
\autoref{tab:result-parameters} shows how different parameters influence the overheads.

We observe that changing the nearby checking range does not introduce observable 
differences for the memory overhead.
The introduced delta is possibly due to server fluctuations.
Using a larger nearby checking range introduces a larger run-time overhead,
as \sys needs to compute and check more canary values.
Using a larger random allocation entropy or RIO introduces 
both larger memory and run-time overheads.

\section{Discussion}
\label{s:discuss}

\PP{Limited UAF-read protection}
\sys provides protection toward UAF-read
based on the assumption that
the attacker cannot distinguish
the memory content stored in the victim data field
from the content stored in other data fields.
However, in the scenario where the victim object
contains a field that the attacker could craft,
e.g., a marker value like \cc{0xdeadbeef},
the victim field can be located
by locating the crafted data field via UAF read(s).

This protection would be strong enough
if the attack is complex enough that
requires multiple successful UAF reads.
However, it should be noted that 
as long as the dangling pointer 
can be used to access the memory,
the attacker can read the data,
and has the chance to identify the offset,
which is the fundamental limitation of all 
statistically secure memory allocators.
For a more comprehensive 
level of protection, 
other measures need to be taken.
Complete mitigation of UAF-read attack is only achieved currently
by either removing dangling pointers or 
pointer nullification.

\PP{Protecting kernel space memory}
While \sys is a user-space memory allocator,
its design is versatile enough
to be adopted for protecting the kernel space.
The Linux kernel operates with a limited number of heap objects,
each field of which is equipped
with finely calibrated permissions.
Consequently,
exploiting the kernel
typically involves a series of intricate steps,
aligns favorably with the
design of \sys.
Should the kernel adopts \sys into its structure,
an attacker would be faced with the daunting task of achieving success
in each individual attack phase,
making a successful attack highly unlikely.
We consider this as future work.

\if 0
For example, 
CVE-2022-32250~\cite{CVE-2022-32250}
is a UAF vulnerability 
in the Linux \cc{Netfilter} subsystem.
One of its PoC attack~\cite{CVE-2022-32250-poc} 
requires the execution of six successful UAF writes to escalate the 
permission. 
This complex attack vector highlights the potential benefits of a \sys-style defense mechanism within the kernel.
We explain one step of its attack in~\autoref{s:cves},
while the attacker needs to successfully launch five 
additional attacks in a row for their desired result.
\fi

\PP{Increasing canary check frequency}
As the core of detecting invalid UAF writes,
\sys checks the nearby canaries when a block is freed.
While this design could virtually cover all memory
slots for a program with frequent heap operations,
UAF attack attempts cannot be detected if there is
no \cc{malloc} call that triggers \sys to check FBC.

While FBC checks could be integrated into other 
heap management functions call logics (e.g., \cc{free}),
the attacker can still possibly bypass the checking 
logic by never calling these functions.
FBC checks need to be integrated into the 
regular program operations to fully address this issue,
while imposing a significant performance overhead.

This problem, then, presents an opportunity for future research. 
A potential area of study could involve the strategic placement of canary checks at selected memory read locations.
Additionally, an additional direction 
could be marking selected memory operations as
sensitive that need to be integrated with FBC checks.
\section{Conclusion}
\label{s:conclusion}

While statistically effective against all
common heap exploitation techniques,
state-of-the-art entropy-based
heap allocators are not tailored
to the \emph{active detection}
of unsuccessful exploitation attempts.
As a result, in reality,
to beat a randomization-based
moving-target scheme,
an attacker can simply launch the same attack,
repeatedly,
potentially with heap spraying,
until success.

In this paper,
we present \sys
to fill the gap
of exploitation attempt detection
without compromising security and performance.
In particular,
we introduce three novel primitives
to the design space of heap allocators:
random in-block offset (RIO),
free block canaries (FBC), and
random bag layout (RBL).
Combined with conventional
BIBOP-style random allocation and heap canaries,
\sys is able to
maintain at least the same level of
protection against
other heap exploitations (e.g., overflows)
and yet still achieves  69\% and 96\% protection rate in two attack scenarios, respectively, 
against UAF exploitation attempts targeting a 64 bytes object,
while only incurs 
marginal performance overhead,
making \sys practical to even production systems.
\section{Acknowledgment}
\label{s:ack}
This work is funded in part by NSERC (RGPIN-2022-03325), the David R. Cheriton endowment, and gifts from Facebook, Amazon, Sui Foundation, and Intel Labs.
We are grateful for the insightful feedback and
guidance from our shepherd,
Dr. R. Sekar which helped us to improve the paper
significantly.

\bibliographystyle{splncs04}
\bibliography{p,sslab,conf}

\appendix
\clearpage
\pagebreak

\section{List of failed tests and corresponding exceptions}
\label{s:failed-tests}

\PP{PARSEC x86}
\begin{itemize}
    \item SlimGuard: ray(mmap error), flu(mmap error), fer(SIGSEGV), ded (SIGSEGV)
\end{itemize}

\PP{SPEC x86}
\begin{itemize}
    \item SlimGuard: 600 (false positive double free), 602(Seg Fault), 623(Seg Fault), 631(Seg Fault), 644(Seg Fault), 657(Seg Fault)
    \item DieHarder: 602(time out), 657(time out)
\end{itemize}

\PP{PARSEC AARCH}
\begin{itemize}
    \item SlimGuard: ray(mmap error), flu(mmap error), fer(SIGSEGV), ded (SIGSEGV)
\end{itemize}

\PP{SPEC AARCH}
\begin{itemize}
    \item SlimGuard: 600 (false positive double free), 602(Seg Fault), 623(Seg Fault), 631(Seg Fault), 644(Seg Fault), 657(Seg Fault)
    \item DieHarder: All benchmarks (Too many open files error)
\end{itemize}

While running SPEC with DieHarder,
test 602 runs unacceptably long, and possibly never returns.
We killed the execution after running 1.5 hours,
and we also mark it as invalid.
As a comparison, running 602 on the AARCH machine with the
default memory allocator only takes about 350 seconds.

NOTE: a PARSEC test is referenced by
the first three letters of its name.

\clearpage
\pagebreak

\if 0
\section{Detailed explanation of how one kernel exploitation can be defended}
\label{s:cves}

\textbf{CVE-2022-32250}~\cite{CVE-2022-32250}
is a UAF vulnerability in the Linux \cc{Netfilter} subsystem.
A dangling pointer can be left
in the \cc{set}$\rightarrow$ \cc{binding} field of the \cc{nft\_lookup} object.
The attacker could reclaim the freed memory
with a \cc{cgroup\_fs\_context} object causing a type confusion attack.
\cc{cgroup\_fs\_context} objects contain a pointer pointing to a memory 
chunk that will be freed.
The attacker can modify the pointed memory address using the dangling pointer
causing arbitrary memory free.

\sys provides UAF read protection in this case with RIO,
statistically guaranteeing that the target field cannot be 
accurately modified using the dangling pointer.

In the PoC attack~\cite{CVE-2022-32250-poc}, 
the attacker must overwrite multiple objects using different
dangling pointers to launch a successful attack,
which is unlikely to be achieved with \sys.
Other probabilistic secure memory allocators do not provide any defense or detection,
since the attacker would spray the memory slots.
Moreover, the memory overwrite happens after the old memory is freed, 
making zeroing out effectless. 

\begin{figure}[ht]
    \centering
    \input{code/example-2}
    \caption{Example UAF attack based on CVE-2022-22620~\cite{cve-2022-22620-poc}}
    \label{code:uaf-example-2}
\end{figure}

\clearpage
\pagebreak
\fi

\section{Run time and memory overheads of each PARSEC and SPEC test}
\label{ss:overheads}

\begin{table}[h]
    \centering
    \begin{tabular}{ccc}
        \toprule
        Setup &
        Run-time overhead &
        Memory overhead \\
        
        \centered{\begin{tabular}[c]{@{}c@{}}PARSEC\\ x86\end{tabular}} & \leftFigure{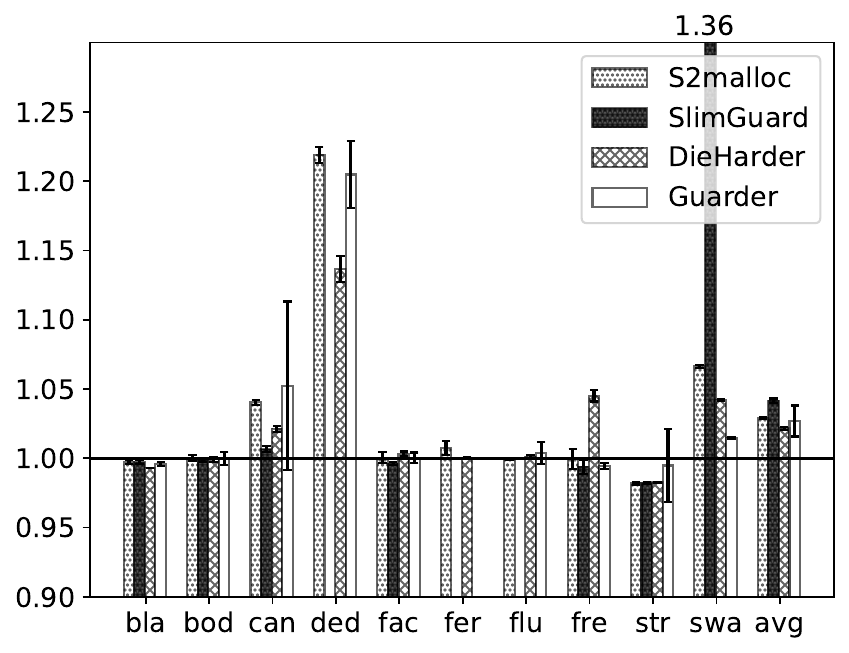} & 
        \rightFigure{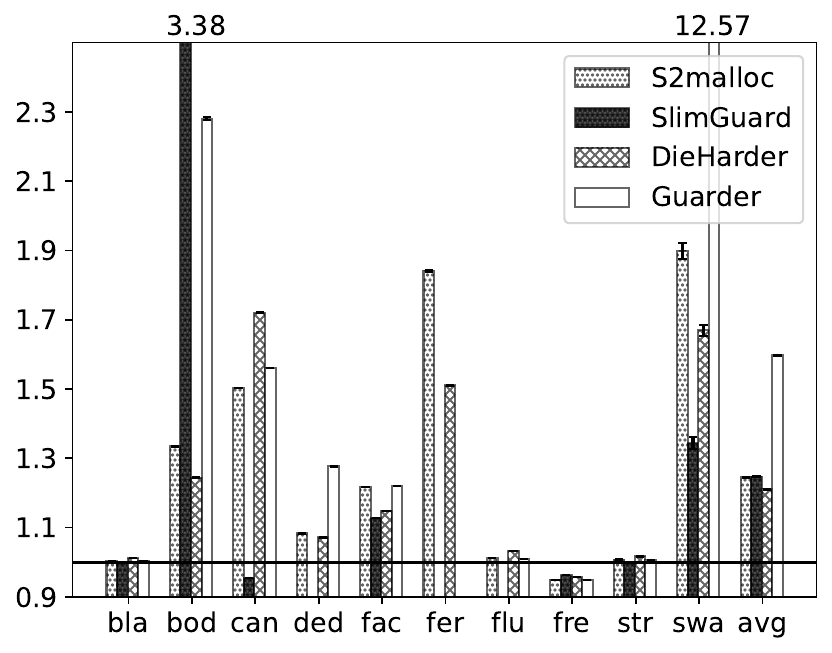} \\
        
        \centered{\begin{tabular}[c]{@{}c@{}}SPEC\\ x86\end{tabular}} & \leftFigure{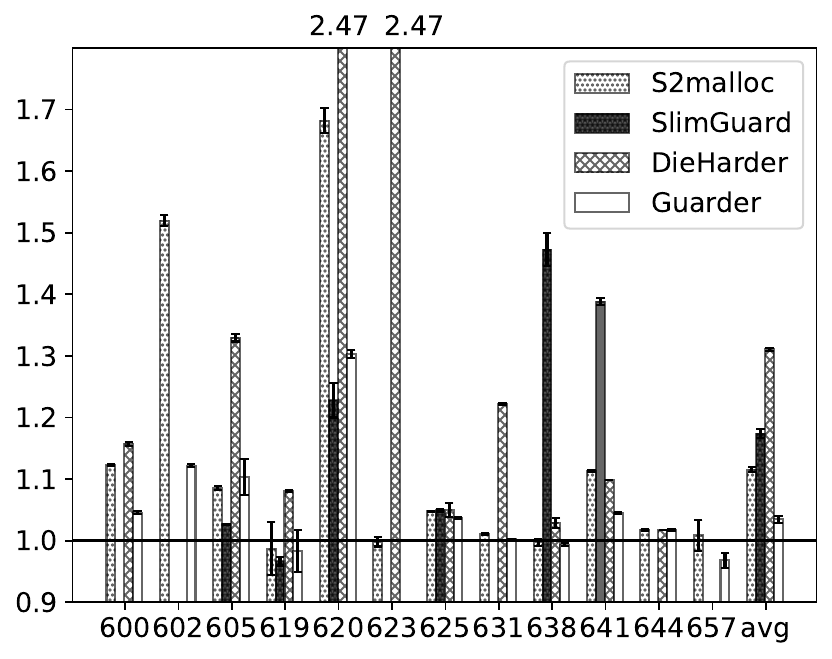} &
        \rightFigure{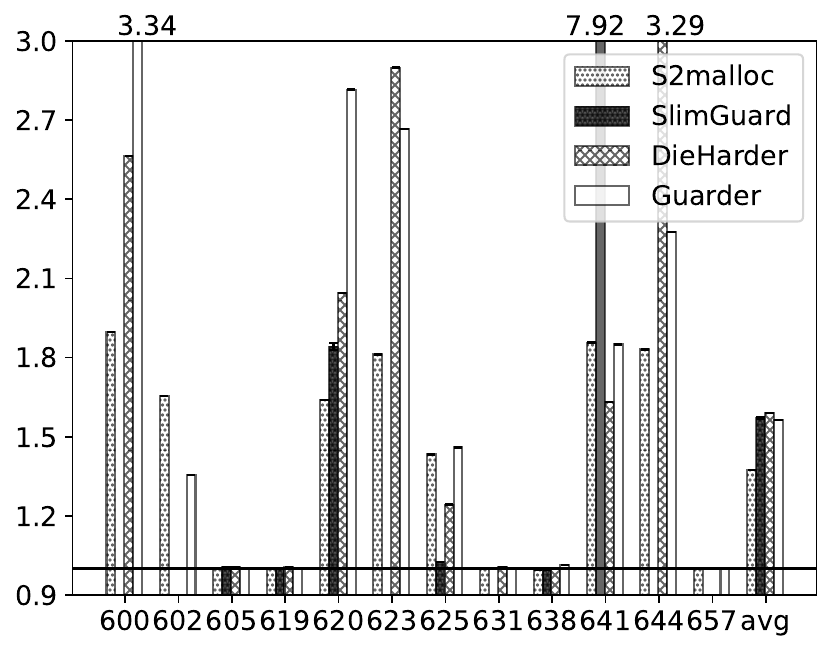} \\
        
        \centered{\begin{tabular}[c]{@{}c@{}}PARSEC\\ AARCH\end{tabular}} & \leftFigure{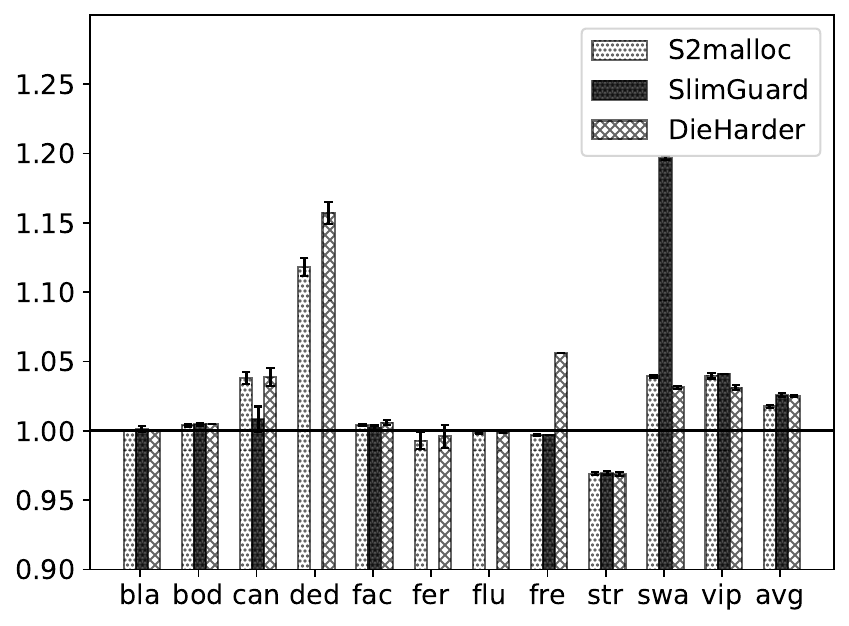} &
        \rightFigure{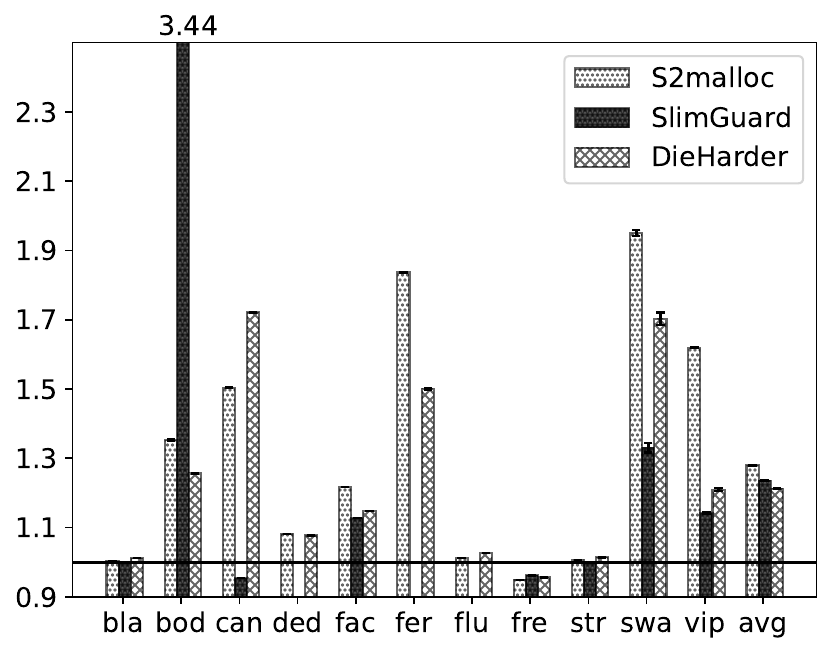} \\
        
        \centered{\begin{tabular}[c]{@{}c@{}}SPEC\\ AARCH\end{tabular}} & \leftFigure{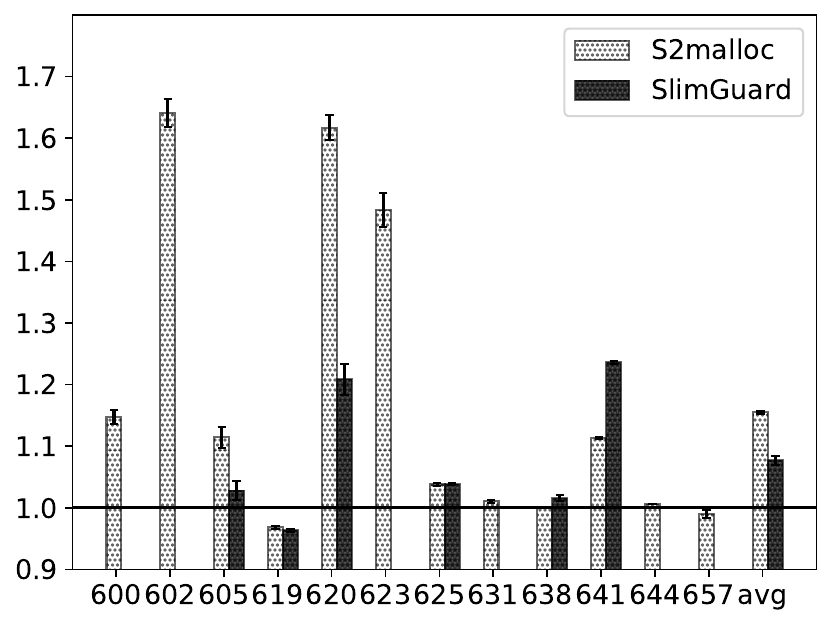} &
        \rightFigure{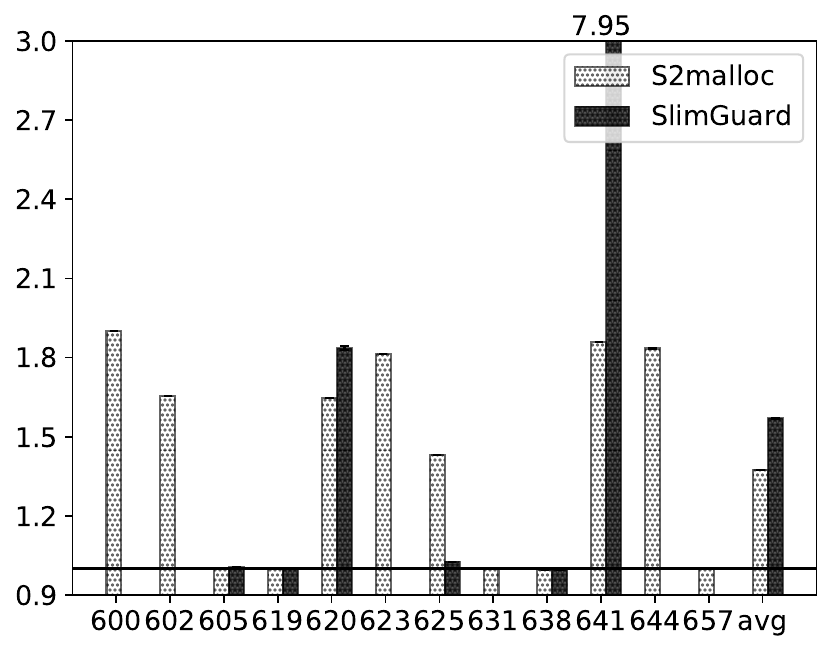} \\
        
        \bottomrule
    \end{tabular}
    \caption{
        Average and standard deviation of run time and memory overhead
        on PARSEC and SPEC benchmarks (x86 and AARCH).
    }
    \label{tab:result-figures}
\end{table}

\clearpage
\pagebreak

\section{Run time and memory overheads of each mimalloc-bench test}
\label{ss:mimalloc-app}

\begin{table}[h]
    \centering
    \begin{tabular}{lcccc}
        \toprule
        & \textbf{\sys} & \textbf{SlimGuard} & \textbf{DieHarder} & \textbf{Guarder} \\
        \midrule
        cfrac & 3.103 & 2.939 & 3.326 & 1.848 \\
        espresso & 2.223 & 6.935 & 2.073 & 1.293 \\
        barnes & 0.999 & 1.000 & 1.024 & 0.998 \\
        leanN & 2.194 & - & 3.297 & 1.498 \\
        larsonN & 9.313 & 23.250 & 7.033 & 1.748 \\
        mstressN & 3.500 & 4.333 & 2.367 & 1.833 \\
        rptestN & 5.383 & 18.131 & 3.243 & 1.212 \\
        gs & 1.599 & 1.616 & 1.333 & 1.184 \\
        lua & 1.579 & 2.364 & 1.337 & 1.200 \\
        alloc-test1 & 3.612 & 3.103 & 4.314 & 1.783 \\
        sh6benchN & 5.392 & - & 40.675 & 2.228 \\
        sh8benchN & 8.384 & - & 26.302 & 3.038 \\
        xmalloc-testN & 2.640 & 4.052 & 5.136 & 1.519 \\
        cache-scratch1 & 1.000 & 1.000 & 1.007 & 1.000 \\
        glibc-simple & 3.370 & 1244.895 & 2.801 & 1.824 \\
        glibc-thread & 7.008 & 14.153 & 4.279 & 2.980 \\
        redis & 1.006 & 1.004 & 1.008 & 0.999 \\
        average & 2.891 & 3.981 & 3.298 & 1.561 \\
        \bottomrule
    \end{tabular}
    \caption{Normalized runtime overheads of mimalloc-bench}
\end{table}

\begin{table}[h]
    \centering
    \begin{tabular}{lcccc}
        \toprule
        & \textbf{\sys} & \textbf{SlimGuard} & \textbf{DieHarder} & \textbf{Guarder}\\
        \midrule
        cfrac & 2.008 & 1.320 & 2.539 & 85.164 \\
        espresso & 23.822 & 386.400 & 3.480 & 99.456 \\
        barnes & 1.034 & 1.014 & 1.115 & 1.045 \\
        leanN & 1.865 & - & 1.900 & 1.988 \\
        larsonN & 11.189 & 5.667 & 0.986 & 17.966 \\
        mstressN & 4.230 & 3.853 & 2.098 & 2.750 \\
        rptestN & 12.056 & 5.119 & 3.224 & 5.661 \\
        gs & 1.548 & 3.218 & 1.748 & 1.683 \\
        lua & 1.760 & 1.353 & 1.333 & 1.494 \\
        alloc-test1 & 2.156 & 1.481 & 1.511 & 28.857 \\
        sh6benchN & 1.103 & - & 1.697 & 3.109 \\
        sh8benchN & 3.430 & - & 1.043 & 8.004 \\
        xmalloc-testN & 580.270 & 8.197 & 3.032 & 1238.115 \\
        cache-scratch1 & 1.461 & 1.250 & 1.564 & 1.422 \\
        glibc-simple & 3.592 & 239.045 & 3.778 & 260.030 \\
        glibc-thread & 15.690 & 3.237 & 3.823 & 75.580 \\
        redis & 1.231 & 1.099 & 1.249 & 1.203 \\
        average & 4.340 & 4.502 & 1.922 & 10.800 \\
        \bottomrule
    \end{tabular}
    \caption{Normalized memory overheads of mimalloc-bench}
\end{table}

\pagebreak
\section{Adapted code snippets to illustrate CVE-2015-6835 and its exploits}
\label{ss:cve-poc}

\begin{figure*}[h]
    \centering
    \tiny
    \hfill\begin{minipage}[c]{.7\linewidth}
    \input{code/example.c}
    \end{minipage}
    \caption{Adapted code snippets to illustrate CVE-2015-6835 and its exploits.}
    \label{fig:cve-poc}
\end{figure*}

\clearpage
\pagebreak

\section{On The Formal Modeling of Probabilistic Use-After-Free Detection}
\label{s:app-formal}

To mathematically model how \sys provides defense against UAF,
we make the following assumptions for the attacker and target program
(which are consistent with our adversary model in~\autoref{ss:design-adv-model}):
\begin{itemize}[noitemsep,topsep=0pt,leftmargin=20pt]
    \item[\WC{1}] The goal of the attacker is to
        modify a sensitive field
        (e.g., a function pointer or an \cc{is_admin} flag)
        in a specific type of object,
        a.k.a., \emph{a victim object},
        via memory writes over a dangling pointer
        (i.e., UAF-writes).

    \item[\WC{2}] The attacker can obtain
        a dangling pointer
        through a bug in the program
        at any point of time during execution.

    \item[\WC{3}] The program
        repetitively allocates and frees
        the type of objects targeted by the attacker
        (i.e., victim objects)
        during its execution.
        However, we do not assume that
        each victim object is freed
        before the next victim is allocated.

    \item[\WC{4}] The attacker can
        either indirectly monitor or directly control
        the allocations of victim objects,
        i.e., the attacker knows when a victim object is allocated,
        but does not know the address of the allocation.

    \item[\WC{5}] Any memory writes
        through the dangling pointer is conducted
        after the victim object is allocated.

    \item[\WC{6}] If the intended sensitive field of a victim object
        is overridden, the attack succeeds;
        otherwise, the program continues to execute,
        allowing the attacker to repeat the exploitation effort
        unless detected by \sys (condition \WC{7}).

    \item[\WC{7}] \sys checks FBCs on each heap allocation and
        detects the attack if any FBC is modified.
\end{itemize}

\noindent
To simplify the illustration,
we assume that the above execution logic
is the only code logic that
involves heap management.
In real-world settings,
attackers usually have an even lower success rate as
memory slots can be allocated to other objects,
which gives \sys more chances to check FBCs and
detect UAF attempts.

\PP{Notation}
We denote the victim object size as $s$
which will be placed in a block of size $b$.
Within a victim object,
the sensitive data starts at the $s_1$ byte,
and with length $l$.
The RIO entropy is $e$ and
obviously, $b \geq e + s$.
The length of the FBC is $c$.
The block-level entropy bit is $n$,
i.e., each allocation of the victim object
will fall in one of $r = 2^n$ blocks.

\subsection{Success rate of attack and defense per single attempt}

In a block hosting a victim object,
the first byte of the sensitive field is
in the interval $[s_1, b - s + s_1)$.
A reasonable attacker will
always try to modify $l$ bytes of data
starting at some byte within the interval.
A smarter attacker will further leverage
the knowledge that
memory allocations are 16 bytes aligned
(a convention from \cc{glibc}).
This implies that
the RIO of a block is randomly chosen from
one out of $1 + \lfloor \dfrac{b-s}{16} \rfloor$ positions.
Thus,
if the attacker attempts to write $l$ bytes
through the dangling pointer
with a randomly guessed RIO value,
the chance of success per trial is:

\begin{equation}
    \tiny
    \begin{aligned}
        A =
        \dfrac{1}{r}
        \vartriangleright\text{\small{the correct block}}\vartriangleleft
        \times
        \dfrac{1}{1 + \lfloor \dfrac{b-s}{16} \rfloor}
        \vartriangleright\text{\small{the correct in-block RIO value}}\vartriangleleft
        = \dfrac{1}{r(1 + \lfloor \dfrac{b-s}{16} \rfloor)}
    \end{aligned}
\end{equation}

\sys puts an FBC randomly to
any $c$ consecutive bytes in the block
with the same probability.
Hence, the probability that
an FBC is modified by an $l$-byte write
in the same block is
reduced to the probability of
selecting
one $l$-byte chunk and one $c$-byte chunk
randomly from a $b$-byte block and
the two chunks overlaps by at least one byte.

\begin{equation}
    \begin{aligned}
        D &=
        \dfrac{
            2\sum_{i=0}^{c-2}(l+i) +
            \sum_{\_={c-1}}^{b-(l+c-1)}(l+c-1)\vartriangleright\text{\small{number of overlaps}}\vartriangleleft
        }{(b-l+1)(b-c+1)\vartriangleright\text{\small{number of ways to place $l$-byte and $c$-byte}}\vartriangleleft}\\ &=
        \dfrac{b(l+c-1) - (l-1)^2 - (c-1)^2 - cl +1}{(b-l+1)(b-c+1)}
    \end{aligned}
\end{equation}

\noindent
The above equation holds when
$b \geq l + 2(c-1)$,
which represents the most practical cases
(i.e., lengths of both sensitive field and FBC are small)
and favors the attacker.
In fact,
if either $l$ or $c$ is large enough relative to $b$,
any $l$-byte write to the block will almost always corrupt the FBC
and can be detected by \sys.

\subsection{Strategy S1: repetitive UAF-writes to the same address}
\label{ss:strategy-1}

In this strategy,
the attacker first
obtains a dangling pointer (\WC{2})
and holds the pointer for arbitrarily long.
Every time the attacker notices a victim object is allocated (\WC{4}),
an $l$-byte UAF-write at the same offset
through the same dangling pointer is conducted (\WC{5}).
This is essentially repetitive UAF-writes to the same address.

As \sys only detects UAF attempts
when a victim object is allocated,
we use round $i$ to represent the $i$-th
allocation of a victim object \emph{after}
the attacker obtains the dangling pointer
and conducts the UAF-write.
Effectively, after round $i$,
\sys should have checked FBCs $i$ times to catch the UAF attempt.

We denote $\mathbb{P}_{e}^{i}$ to represent
the probability that the program execution ever reaches round $i$.
By definition,
$\mathbb{P}_{e}^{1} = 1 - A$,
i.e., when the attacker's first UAF-write is not successful
in achieving the goal (\WC{1}).
Suppose the execution has reached round $i$,
the probability that the repetitive UAF-writes
is detected at this particular round is

\begin{equation}
    \begin{aligned}
        \dfrac{2d+1}{r} &\vartriangleright\text{\small{the FBC of the overridden block is checked}}\vartriangleleft
        \times \\D &\vartriangleright\text{\small{the FBC of the overridden block is corrupted}}\vartriangleleft
    \end{aligned}
    \label{e:s1-detected}
\end{equation}

\noindent
Based on this,
we can derive the inductive definition for $\mathbb{P}_{e}^{i}$:

\begin{equation}
    \begin{aligned}
        \mathbb{P}_{e}^{i+1} =
        \mathbb{P}_{e}^{i} &\vartriangleright\text{\small{reaches round $i$}}\vartriangleleft
        \times (1 - \dfrac{2d+1}{r} \times D) \vartriangleright\text{\small{undetected}}\vartriangleleft
        \\ \times (1 - A) &\vartriangleright\text{\small{unsuccessful attack attempt}}\vartriangleleft
    \end{aligned}
\end{equation}

\noindent
Limiting program execution to an upper bound of $K$ rounds,
the chance of attacker and \sys wins, respectively, is:

\begin{equation}
    \begin{aligned}
        \mathbb{P}_{attack}^{K} = A + \sum_{i=1}^{K} (\mathbb{P}_{e}^{i} \times A)
        \quad\quad :: \quad\quad
        \mathbb{P}_{detect}^{K} = \sum_{i=1}^{K} (\mathbb{P}_{e}^{i} \times \dfrac{2d+1}{r} \times D)
    \end{aligned}
\end{equation}

\subsection{Strategy S2: UAF-writes through fresh dangling pointers}
\label{ss:strategy-2}

Unlike~\autoref{ss:strategy-1},
the attacker does not hold a dangling pointer indefinitely,
instead, the attacker obtains a fresh dangling pointer (\WC{2})
if a prior UAF-write attempt is not successful.
After obtaining a dangling pointer,
if the attacker notices a victim object is allocated (\WC{4}),
an $l$-byte UAF-write through the fresh dangling pointer
is conducted (\WC{5}).
This essentially means that
every UAF-write is likely on a different address.

More importantly,
as \sys only detects UAF attempts
when a victim object is allocated,
this strategy effectively creates
a turn-based game between the attacker and \sys
where in each round,
the attacker makes the move of
obtaining a dangling pointer and conducting a UAF-write
while
\sys makes the move of checking FBCs and allocating a new victim object
(if FBCs checked are intact).
The game ends when either the attacker or \sys wins.

We use round $i$ to represent the $i$-th
round of the game.
In each round, the attacker makes the first move and \sys follows.
We denote $\mathbb{P}_{e}^{i}$ to represent
the probability that the program execution ever reaches
\sys's turn in round $i$
(to be consistent with the notation
in~\autoref{ss:strategy-1}).
By definition,
$\mathbb{P}_{e}^{1} = 1 - A$,
i.e., the attacker's first UAF-write is not successful
in achieving the goal (\WC{1}).

To calculate the detection rate by \sys,
we rephrase the question to a classical combinatorics question:
there exists $r$ balls in a box where
each time the attacker
picks one ball randomly (i.e., the block referred by the dangling pointer),
colors it with probability $D$ (i.e., corrupts the FBC in the block), and
puts the ball back to the box.
A ball cannot be uncolored once it is colored
(because the attacker does not undo a UAF-write).
We use $\mathbb{Q}_i$ to denote
the probability that an arbitrary ball in the box is not colored
(i.e., a block with its FBCs integral)
after $i$ rounds.

\begin{equation}
\tiny
    \begin{aligned}
        \mathbb{Q}_i =
        (\dfrac{r-1}{r} \vartriangleright\text{\small{ball not selected}}\vartriangleleft +
        \dfrac{1}{r} \times (1-D) \vartriangleright\text{\small{ball selected but not colored}}\vartriangleleft)^{i}
        = (\dfrac{r-D}{r})^{i}
    \end{aligned}
\end{equation}

\noindent
Therefore, at round $i$,
there will be, by expected value,
$r \cdot Q_i$ balls remain uncolored in the box.
The detection rate of \sys at round $i$ will be
the same as the probability of
selecting $2d+1$ \emph{consecutive} balls
from a string of $r$ balls
where at least one of the selected balls is colored.
The detection rate
is denoted as $\mathbb{P}_{d}^{i}$ and calculated as:

\begin{equation}
    \begin{aligned}
        \mathbb{P}_{d}^{i} = 1 - \dfrac{Q_ir - 2d}{r - 2d} \cdot \prod_{i=0}^{2d}(\dfrac{Q_ir-i}{r-i})
    \end{aligned}
\end{equation}

\noindent
Based on this,
we can derive the inductive definition for $\mathbb{P}_{e}^{i}$:

\begin{equation}
    \begin{aligned}
        \mathbb{P}_{e}^{i+1} =
        \mathbb{P}_{e}^{i} &\vartriangleright\text{\small{reaches round $i$}}\vartriangleleft
        \times (1 - \mathbb{P}_{d}^{i}) \vartriangleright\text{\small{undetected}}\vartriangleleft
        \times \\ (1 - A) &\vartriangleright\text{\small{unsuccessful attack attempt}}\vartriangleleft
    \end{aligned}
\end{equation}

\noindent
Limiting program execution to an upper bound of $K$ rounds,
the chance of attacker and \sys wins, respectively, is:

\begin{equation}
    \begin{aligned}
        \mathbb{P}_{attack}^{K} = A + \sum_{i=1}^{K} (\mathbb{P}_{e}^{i} \times A)
        \quad\quad :: \quad\quad
        \mathbb{P}_{detect}^{K} = \sum_{i=1}^{K} (\mathbb{P}_{e}^{i} \times \mathbb{P}_{d}^{i})
    \end{aligned}
\end{equation}

\subsection{Strategy S1-spray: repetitive UAF-writes to the same address with spraying}
\label{ss:strategy-3}

This strategy operates similarly
to the strategy in~\autoref{ss:strategy-1}:
the attacker first
obtains a dangling pointer (\WC{2})
and holds the pointer for arbitrarily long.
However, generalized from~\autoref{ss:strategy-1}
in which the attacker conducts a UAF-write
after one allocation of a victim object (\WC{4}),
the attacker waits until there are $m$
victim objects newly allocated and alive (i.e., not freed yet)
and then issues the UAF-write.
Effectively,
the attacker is trying to diligently spray the heap by
victim objects to increase its chance of success.
The UAF-write is still an $l$-byte memory write at the same offset
through the same dangling pointe (\WC{5}).
This is essentially repetitive UAF-writes to the same address,
similar to~\autoref{ss:strategy-1}.

Consistent with the analysis in~\autoref{ss:strategy-1},
we still use round $i$ to represent the $i$-th
allocation of a victim object \emph{after}
the attacker obtains the dangling pointer
and conducts the UAF-write.
Effectively, after round $i$,
\sys should have checked FBCs $i$ times to catch the UAF attempt.

We denote $\mathbb{P}_{e}^{i}$ to represent
the probability that the program execution ever reaches round $i$.
By definition,
$\mathbb{P}_{e}^{1} = 1 - mA$,
i.e., when the attacker's first UAF-write is not successful
in achieving the goal (\WC{1}).
Note that the attacker success rate increases
as there are $m$ victim objects alive and
the attack succeeds
as long as the sensitive field in
any one of them is overridden by the UAF-write---this
is essentially the advantage of heap spraying.

And yet, consistent with~\autoref{ss:strategy-1},
this UAF-write can corrupt one FBC at most.
Hence,
suppose the execution has reached round $i$,
the probability that the repetitive UAF-writes
is detected at this particular round is
still~\autoref{e:s1-detected}.
With attacker's success rate improved,
the inductive definition for $\mathbb{P}_{e}^{i}$
in this strategy will be:

\begin{equation}
    \begin{aligned}
        \mathbb{P}_{e}^{i+1} =
        \mathbb{P}_{e}^{i} &\vartriangleright\text{\small{reaches round $i$}}\vartriangleleft
        \times (1 - \dfrac{2d+1}{r} \times D) \vartriangleright\text{\small{undetected}}\vartriangleleft
         \times \\(1 - mA) &\vartriangleright\text{\small{unsuccessful attack attempt}}\vartriangleleft
    \end{aligned}
\end{equation}

\noindent
Limiting program execution to an upper bound of $K$ rounds,
the chance of attacker and \sys wins, respectively, is:

\begin{equation}
    \begin{aligned}
        \mathbb{P}_{attack}^{K} = mA + \sum_{i=1}^{K} (\mathbb{P}_{e}^{i} \times mA)
        \quad\quad :: \quad\quad
        \mathbb{P}_{detect}^{K} = \sum_{i=1}^{K} (\mathbb{P}_{e}^{i} \times \dfrac{2d+1}{r} \times D)
    \end{aligned}
\end{equation}

\subsection{Strategy S2-spray: UAF-writes through fresh dangling pointers with spraying}
\label{ss:strategy-4}

In this strategy,
the attacker still sprays the heap
such that overriding the sensitive field
in any the $m$ victim objects achieves the goal
(like the strategy in~\autoref{ss:strategy-3}).
Similar to~\autoref{ss:strategy-2},
the attacker does not hold a dangling pointer indefinitely,
instead, the attacker obtains a fresh dangling pointer (\WC{2})
if a prior UAF-write attempt is not successful and use it to launch a UAF-write attack in the current round.
In each round, the attacker makes the first move and \sys follows.
We denote $\mathbb{P}_{e}^{i}$ to represent
the probability that the program execution ever reaches
\sys's turn in round $i$
(to be consistent with the notation
in~\autoref{ss:strategy-1} ,~\autoref{ss:strategy-2},
and~\autoref{ss:strategy-3}).
By definition,
$\mathbb{P}_{e}^{1} = 1 - mA$,
i.e., the attacker's first UAF-write is not successful
in achieving the goal (\WC{1}) even after spraying $m$ victim objects.

Similar to~\autoref{ss:strategy-2},
we use $\mathbb{Q}_i$ to denote
the probability that an arbitrary ball in the box is not colored
(i.e., a block with its FBCs integral)
after $i$ rounds.
$\mathbb{Q}_i$ and $\mathbb{P}_{d}^{i}$ can be computed using the same formula as in~\autoref{ss:strategy-2}

Based on this,
we can derive the inductive definition for $\mathbb{P}_{e}^{i}$:

\begin{equation}
    \begin{aligned}
        \mathbb{P}_{e}^{i+1} =
        \mathbb{P}_{e}^{i} &\vartriangleright\text{\small{reaches round $i$}}\vartriangleleft
        \times (1 - \mathbb{P}_{d}^{i}) \vartriangleright\text{\small{undetected}}\vartriangleleft
        \times\\ (1 - mA) &\vartriangleright\text{\small{unsuccessful attack attempt}}\vartriangleleft
    \end{aligned}
\end{equation}

\noindent
Limiting program execution to an upper bound of $K$ rounds,
the chance of attacker and \sys wins, respectively, is:

\begin{equation}
    \begin{aligned}
        \mathbb{P}_{attack}^{K} = mA + \sum_{i=1}^{K} (\mathbb{P}_{e}^{i} \times mA)
        \quad\quad :: \quad\quad
        \mathbb{P}_{detect}^{K} = \sum_{i=1}^{K} (\mathbb{P}_{e}^{i} \times \mathbb{P}_{d}^{i})
    \end{aligned}
\end{equation}

\if 0
\subsection{Strategy 4: UAF-writes through fresh dangling pointers with reset}
\label{ss:strategy-4}

This strategy operates similarly
to the strategy in~\autoref{ss:strategy-2}:
the attacker does not hold a dangling pointer indefinitely,
instead, the attacker obtains a fresh dangling pointer (\WC{2})
if a prior UAF-write attempt fails.

However, different from~\autoref{ss:strategy-2},
the attacker will diligently reset the previous UAF-write
before releasing the old dangling pointer.
A UAF-write can be reset
if the attacker first UAF-read
the $l$-byte chunk before the write---the
reset is simply overriding the UAF-write
by the value from the UAF-read.
In this way,
if the UAF-write accidentally modifies an FBC,
the reset will effectively restore the FBC.

What happens
after obtaining a dangling pointer is the same as~\autoref{ss:strategy-2}:
if the attacker notices a victim object is allocated (\WC{4}),
an $l$-byte UAF-write through the fresh dangling pointer
is conducted (\WC{5}).
This essentially means that
every UAF-write is likely on a different address.
\fi

\end{document}